\documentclass[reprint,
superscriptaddress,
 amsmath,amssymb,
 aip,jcp,longbibliography
]{revtex4-2}

\usepackage{graphicx}
\usepackage{dcolumn}
\usepackage{bm}
\usepackage{lipsum}
\usepackage{xcolor}
\usepackage{mathrsfs}
\usepackage{multirow}
\usepackage{tabularray}

\begin{document}

\title{Vibrational strong coupling influences product selectivity in a model for post-transition-state bifurcation reactions}

\author{Subhadip Mondal}
\altaffiliation[Current address: ]{Department of Chemistry, University of Rochester, Rochester, NY 14627, USA}
\affiliation{Department of Chemistry,
 Indian Institute of Technology, Kanpur, Uttar Pradesh 208 016, India}

\author{Atul Kumar}
\affiliation{Department of Chemistry,
 Indian Institute of Technology, Kanpur, Uttar Pradesh 208 016, India}

\author{Srihari Keshavamurthy}%
\email{srihari@iitk.ac.in}
\affiliation{Department of Chemistry,
 Indian Institute of Technology, Kanpur, Uttar Pradesh 208 016, India}


\begin{abstract}
In this study we explore the possibility of modulating product selectivity (branching ratios) in post transition state bifurcation (PTSB) reactions via vibrational strong coupling (VSC) to an optical cavity. Detailed classical and quantum dynamical calculations on a model potential reveal that the branching ratio can be enhanced  by nearly a factor of two under VSC conditions. Interestingly,  upon altering the shape of the potential, we find a switch in the cavity frequency at which  maximum enhancement in selectivity is observed. Apart from emphasizing the role of both cavity-system and intramolecular energy transfer to the observed enhancements, we highlight the complexity of the VSC mechanism in terms of the choice of the cavity frequency vis-à-vis the various molecular mode frequencies.  Our work shows that, in principle, cavity quantum electrodynamics can reshape dynamical outcomes in reactions with complex potential energy landscapes. 
\end{abstract}

\maketitle

\section{Introduction}
\label{introduction}

 Recently, an approach towards achieving the ``holy grail" of mode-specific chemistry has led to considerable enthusiasm.  In this vibrational strong coupling (VSC) approach, molecules are placed inside an optical cavity and the cavity mode is tuned to resonance with a specific molecular vibration of interest\cite{thomas2016ground,thomas2019tilting,nagarajan2021chemistry,ahn2023modification,Kenji2020,lather2019cavity,mahatoexploring}. The resulting formation of the hybrid light-matter states (polaritons) is expected to have a significant impact on rates and mechanisms of various reactions even in the absence of external light. However, it is important to mention that there have been ``null" experiments\cite{FidlerJCP2023,ChenNanophot2024,mullerdiels} which challenge our current understanding of the VSC mechanism.   In addition, issues regarding experimental reproducibility\cite{imperatore2021reproducibility,wiesehan2021negligible} and possible optical artifacts\cite{michon2024impact,veraguwearxiv2026,michon2026systematic} that can lead to spurious effects have been raised. Consequently, there is an ongoing discussion regarding identifying  effects on reaction rates that are purely due to VSC, genuinely quantum\cite{lindoy2023quantum,ke2024quantum,fiechter2023quantum,lieberherrJCP2023} and beyond the optical filtering regime\cite{schwennicke2025molecular}. Nevertheless, there has been substantial progress\cite{li2021collective,schafer2022shining,chen2022cavity,li2022qm,mondal2022dissociation,gomez2023vibrational,mondal2023phase,yu2023manipulating,mondal2024cavity,yu2025theoretical,ying2024resonance,schafer2024machine,GarnerJCP2025,ji2025selective}  towards gaining insights into the possible mechanism of VSC in terms of the crucial role played by the cavity-orchestrated molecular energy transfer process.
 Apart from reaction rate modulation, the influence of VSC has been investigated in a wide variety of systems. Recent examples include enhancement of electrical conductance in nonconducting polymers\cite{kumar2024extraordinary}, quantum sensing\cite{zheng2025quantum,xiang2024molecular}, modification of glass transition temperature of polymer films\cite{dang2025altering}, and fabrication of nanoantennas\cite{SepulvedaJPCC2025}. 

Most of the theoretical studies of VSC-modulated reaction rates mentioned above have employed models involving a single reactant and product separated by a transition state. Although this is a  useful minimal model in the context of investigating VSC effects on reaction rates,  multidimensional potential energy surfaces (PES) can harbor features like bifurcating reaction paths, higher rank saddles, multiple transition states from a single reactant region, and shallow intermediate states\cite{Minyaev1994883,MaedaIJQC2015}. Such features invariably lead to the existence of competing mechanisms and nonstatistical reaction dynamics. Moreover, the multiple critical points on such multidimensional PES present a choice of a large number of molecular frequencies to which the cavity can be coupled resonantly. An early example for preferential selectivity under VSC comes from the experiment of Thomas et al\cite{thomas2019tilting} wherein it was shown that the reactant  Si--O bond cleavage occurs in favor of the reactant Si--C bond cleavage. Intriguingly, although the Si--C cleavage is  kinetically favored in off-resonant conditions, the Si--O bond cleavage is favored irrespective of whether the cavity resonantly couples with the Si--C or the Si--O modes. Qualitatively similar observations were theoretically obtained\cite{li2021theory} by Li, Mandal, and Huo on a model system by invoking the concept of cavity-induced dynamic caging arising due to the differing barrier curvatures near the two transition states. However, to the best of our knowledge, there is no  clear explanation for the original experimental\cite{thomas2019tilting} observations.  

A key question in the context of VSC is whether or not one can  predict the specific molecular frequencies that would lead to the strongest modulation of the rates, mechanisms, product yields, or selectivity. Answering the question is challenging since this requires insights into the mechanism by which VSC perturbs the cavity-free reaction dynamics and intramolecular vibrational energy redistribution (IVR)\cite{uzermillerphysrep,LehmannARPC1994,nesbitt1996vibrational} pathways. This is particularly true in the case of reactions that involve several competing pathways and multiple transition states leading to different products -- a scenario that is expected to be typical in multidimensional PESs and gives rise to strong dynamical effects. Consequently, given the advances\cite{gruebele2004vibrational,leitnerlrmt2015,karmakar2020intramolecular} in our understanding of intramolecular energy flow dynamics and the ways\cite{LeeJPCL2012,WindhornJCP2003,maitrajpcl2021,shaon2025tuning,hassani2025modulating,SchmitzJPCA2019,rather2019dinitrogen} to manipulate it,  one hopes that detailed dynamical studies on appropriate model systems will lead to valuable insights into the mechanism of VSC. 

\begin{figure}[tbp]
    \begin{center}
    \includegraphics[width=8.0cm]{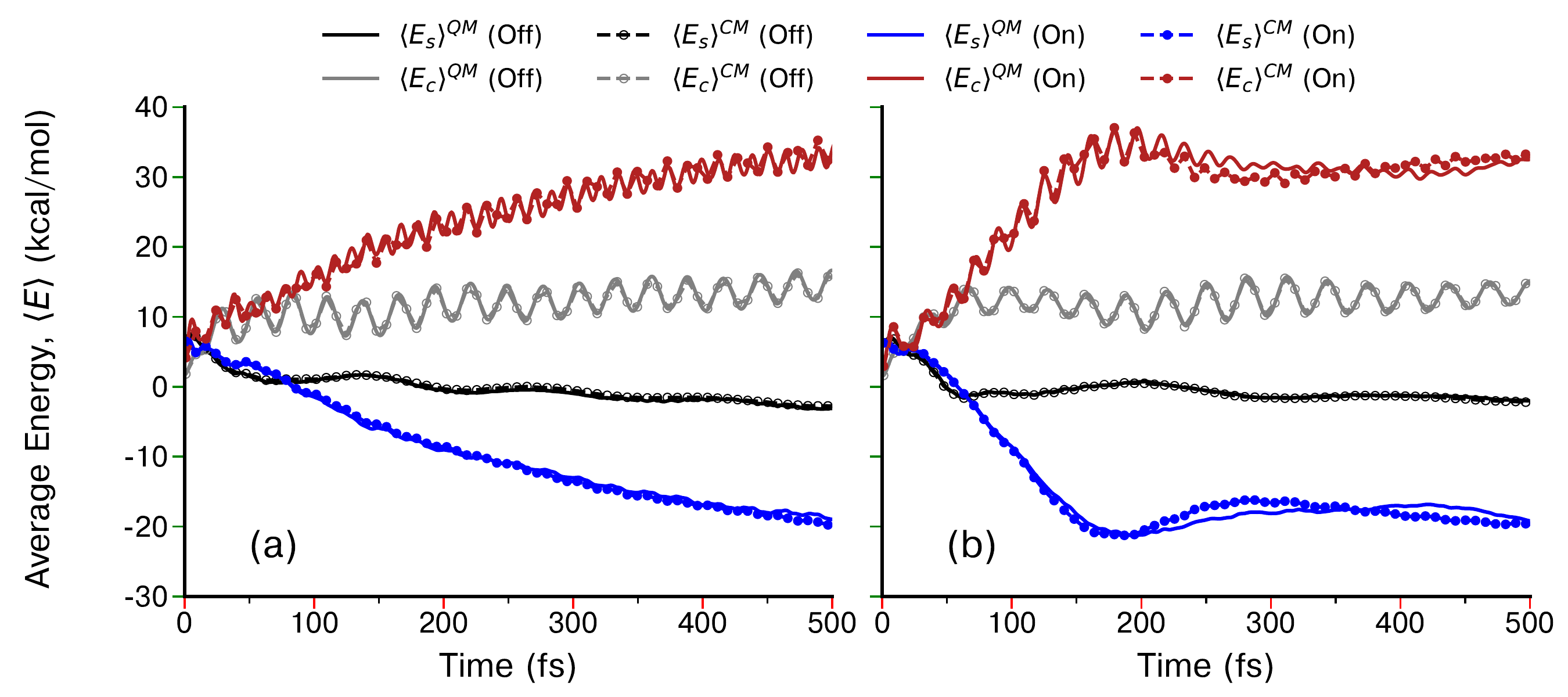}
    \caption{Model potential energy surface Eq.~\ref{modelPES}  exhibiting a post transition state bifurcation. The reactant (R), transition states (TS$_{1}$, TS$_{2}$), products (P$_{1}$, P$_{2}$), and the valley ridge inflection (VRI) point are indicated.}
    \label{fig:PES}
    \end{center}
    \end{figure}

In this context, a good example model system is concerned with  reactions that exhibit a post transition state bifurcation (PTSB)\cite{caramella2002unexpected} due to the presence of a valley ridge inflection point on the PES. Essentially, as shown in Fig.~\ref{fig:PES}, PTSB involves a single entrance transition state (also called ``ambimodal")  bifurcating into two possible products without traversing through any intermediate states.  Notice that the PES shown in Fig.~\ref{fig:PES} is different from the model that one invokes in the context of the earlier work\cite{thomas2019tilting,li2021theory}. The PES model appropriate to the earlier work\cite{thomas2019tilting} involves two different transition states leading to two different products. In particular, while such model PESs can be invoked even in one dimension, the PTSB topology requires a minimum of two dimensions. In this work we study the PTSB case since an increasing number of reactions, ranging from cycloadditions to biosynthetic\cite{HussainNPR2026}, exhibit PTSB and have attracted considerable attention\cite{ess2008bifurcations,hare2017post,rehbein2011we,rehbein2015chemistry,LamChemSci2023,bharadwaz2021bifurcating} due to the fact that in such systems dynamical effects\cite{collins2013nonstatistical,kumar2025quantum,garcia2021dynamical} and the shape of the PES\cite{tao2023symmetric}, instead of the transition state energetics, control the product selectivity.  In fact, a very recent perspective\cite{NielsenJACS2026} highlights the necessity of  dynamical viewpoints for PTSB reactions. 
Thus, the PTSB system provides us with the various desirable properties mentioned above in the context of VSC modulated selectivity. It has multiple critical points, with their associated frequencies, to resonantly couple with the optical cavity mode and presents the simplest model for dynamically determined product selectivity.
More importantly, \textit{ab initio} dynamical studies\cite{bogle2012dynamic,EssJPCA2020}  of  PTSB systems have emphasized the role of specific vibrational modes that strongly influence the selectivity. Thus, despite the fact that VSC experiments are yet to be performed on PTSB reactions, our work makes specific predictions regarding the effect of VSC in such systems in terms of  selectively manipulating the branching ratio by coupling the cavity mode to such special molecular vibrational modes.

Here, we show that VSC does influence the product selectivity in the model system with clear indications that the cavity-free dynamical effects can be amplified in the strong coupling regime. We find that, both classically and quantum mechanically, VSC with the fundamental transition of the reactant well has minimal effect on the selectivity, whereas VSC involving the product wells leads to substantial influence on the selectivity. This observation,  yet to be verified experimentally, resonates with the findings in a recent different model theoretical study by Ke\cite{ke2025non} wherein maximal rate enhancement was obtained at a cavity frequency corresponding to the product well transition, and different from the peak of the molecular absorption spectrum.   However,  despite being a single molecule and single cavity mode study, we note that the work by Ke includes cavity losses and a solvent bath, while our model presented here does not. Furthermore, as a surprising result in this context, we find that altering the depth of the product wells leads to a shift in the cavity frequency at which the maximum effect on the selectivity is observed. The mechanism involves a combination of factors involving cavity-system resonant energy exchange and interwell transitions leading to a cavity-mediated coherent cooling into both the product wells. The differing extent of cooling in the two product wells gives rise to the modulation of the branching ratio.  

We mention at the outset that since our model involves a single cavity mode and a single molecule with no losses or dissipation, our study is in a regime far removed from typical VSC experiments which  involve $\sim 10^{8}$ molecules. The conclusions section provides further discussions on this aspect. However, given the increasing interest in the few molecule strong coupling regime\cite{Creddynat2016,Liunatcomms2024,Robledoarxiv2026,MelliniJCP2026,koner2023path}, our study may yet provide useful insights. 

We start with a description of the model Hamiltonian in Sec.~\ref{modelH} followed by our choice of the initial state and dipole in Sec.~\ref{choicemupsi0}. In Sec.~\ref{bratio} we provide details regarding the definition and computation of the branching ratio for the model system. The results and discussions are given in Sec.~\ref{results} along with possible mechanistic origins of the observed effects.  Section~\ref{conclusion} concludes with brief remarks on the robustness of the single molecule observations in the collective limit and further studies that are necessary for a complete understanding of the VSC mechanism in PTSB systems.

\section{Theoretical preliminaries}
\label{theorprelims}

\subsection{Model Hamiltonian}
\label{modelH}

    The model Hamiltonian studied here, in the Coulomb gauge with the dipole approximation, has the Pauli-Fierz form and can be expressed as
    \begin{equation}
        H({\mathbf r},{\mathbf p},q_{c},p_{c}) = H_{M}({\mathbf r},{\mathbf p}) + \frac{1}{2} p_{c}^{2} + \frac{1}{2} \omega_{c}^{2}\left[q_{c} + \frac{\lambda_{c}}{\omega_{c}} \mu({\mathbf r})\right]^{2}
        \label{modvscham}
    \end{equation}
    where ${\mathbf r} \equiv (x,y)$ and ${\mathbf p} \equiv (p_{x},p_{y})$ correspond to the matter coordinates and their conjugate momenta, respectively. The photon coordinate and momentum $(q_{c},p_{c})$ are associated with a single-mode cavity field of frequency $\omega_{c}$. The cavity field and matter are coupled via the dipole $\mu({\mathbf r})$ with strength $\lambda_{c}$, and note that the form of the interaction used includes the dipole self-energy term.  We have taken the dipole to be aligned with the field polarization, as is typically assumed in such model studies. 
    The matter Hamiltonian is given by  
    \begin{equation}
        H_{M}({\mathbf r},{\mathbf p}) = \frac{1}{2m}(p_{x}^{2} + p_{y}^{2}) + U({\mathbf r})
    \end{equation}
    with $m$ being the mass and the two-dimensional potential energy surface (PES)\cite{tao2023symmetric,garcia2021dynamical}
     
    \begin{equation}
    \begin{split}
    U(\mathbf r) &= \overline{V}^{\ddagger}\left[Ax^{4}- Bx_sx^{3}- Cx_s^{2}x^{2}\right] + Dy^{2}+ Fy^{4} \\
    &\quad + Gx^{2}y^{2}+ Hxy^{2}(y^{2}-2)+ \alpha xy
    \end{split}
    \label{modelPES}
    \end{equation}
      models reactions which exhibit post transition–state bifurcation (PTSB). The from of the PES is shown in Fig.~\ref{fig:PES} where the reactant (R), products (P$_{1}$, P$_{2}$), and the two transition states (TS$_{1}$, TS$_{2}$) are indicated. 

\begin{figure}[tbp]
    \begin{center}
    \includegraphics[width=8.6cm,height=9.8cm]{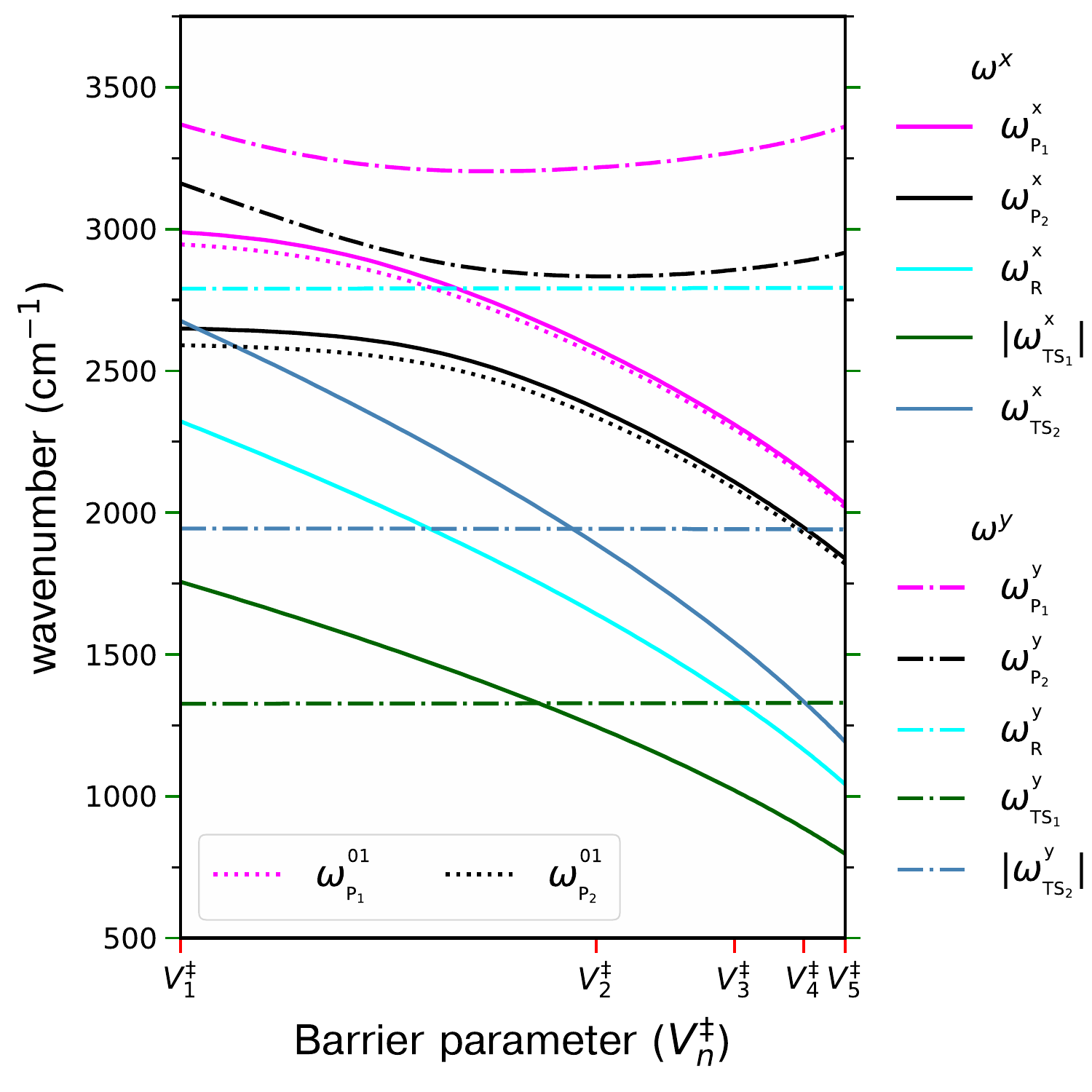}
    \caption{Variation of the harmonic frequencies associated with the different stationary points (indicated) with the parameter $V^{\ddagger}_{n}$. Solid and dash-dot lines show the $x$-mode and the $y$-mode frequencies respectively.  In addition, dotted lines show the $x$-mode fundamental frequencies ($0 \rightarrow 1$) associated with the two product wells.}
    \label{fig:harmfreq_variationwithVdag}
    \end{center}
    \end{figure}

     The PES depends on several parameters that determine the overall shape of the PES and the locations of the various critical points.  The parameter $V^{\ddagger}$ determines the reactant well depth relative to TS$_{1}$ and the relative depths of the product wells.   The parameter $x_{s}$ sets the distance between R and TS$_{2}$ while the parameters $(\alpha,G,H)$ tune the coupling between the molecular modes. In particular, $\alpha$ controls the extent of energetic asymmetry of the two product wells with $\alpha=0$ corresponding to the symmetric case. We fix the parameter values (in au), adapted from previous works\cite{tao2023symmetric,kumar2025quantum}, to be  $V^{\ddagger}= 3/1280$, $x_{s}=0.2$, $A=G=1/2$, $B=2/3$, $C=12$, $D=F=2/5$, $H = 1$, $\alpha=0.1$, and mass $m = 22034$ (mass of carbon atom).  The choice for the value of the mass ensures that, given the values of the rest of the PES parameters, the various vibrational mode frequencies are in the mid-IR range (see Fig.~\ref{fig:harmfreq_variationwithVdag}). Moreover, in typical reactions\cite{kramer2015reaction,hare2017post} exhibiting PTSB, the key collective motions involve heavy atom vibrations and hence the effective mass is expected to be large.   Further details about the location, energies, and stability of the critical points are provided in the Supplementary Sec. S$1$.

    Since the present study is focused on the possible influence of VSC on the selectivity between P$_{1}$ and P$_{2}$, the parameter $\overline{V}^{\ddagger} \equiv V^{\ddagger}_n/x_{s}^{4}$ with $V^{\ddagger}_n \equiv V^{\ddagger}/n$ is of particular interest. This is due to the fact that changing the value of $n$ alters the harmonic frequencies, as shown in Fig.~\ref{fig:harmfreq_variationwithVdag}, of the various critical points without changing the locations of the reactant well R and the transition states TS$_{1}$ and TS$_{2}$.  Clearly, Fig.~\ref{fig:harmfreq_variationwithVdag} shows that decreasing $V^{\ddagger}_{n}$ (increasing $n$) leads to decreasing  $x$-mode frequencies of all the critical points, while the $y$-mode frequencies  remain nearly constant. Thus, in the context of VSC, varying $V^{\ddagger}_{n}$ allows one to systematically explore the conditions under which enhancement, suppression, as well as  ``null" effect on the observable of interest might occur.  Thus, the Hamiltonian in Eq.~\ref{modvscham} provides a minimal model which enables us to probe how the PES topology and cavity-molecule interactions jointly govern the selectivity.

\subsection{Choice of the dipole function and the initial state}
\label{choicemupsi0}

\begin{figure}[t]
    \begin{center}
    \includegraphics[width=8.6cm,height=12cm]{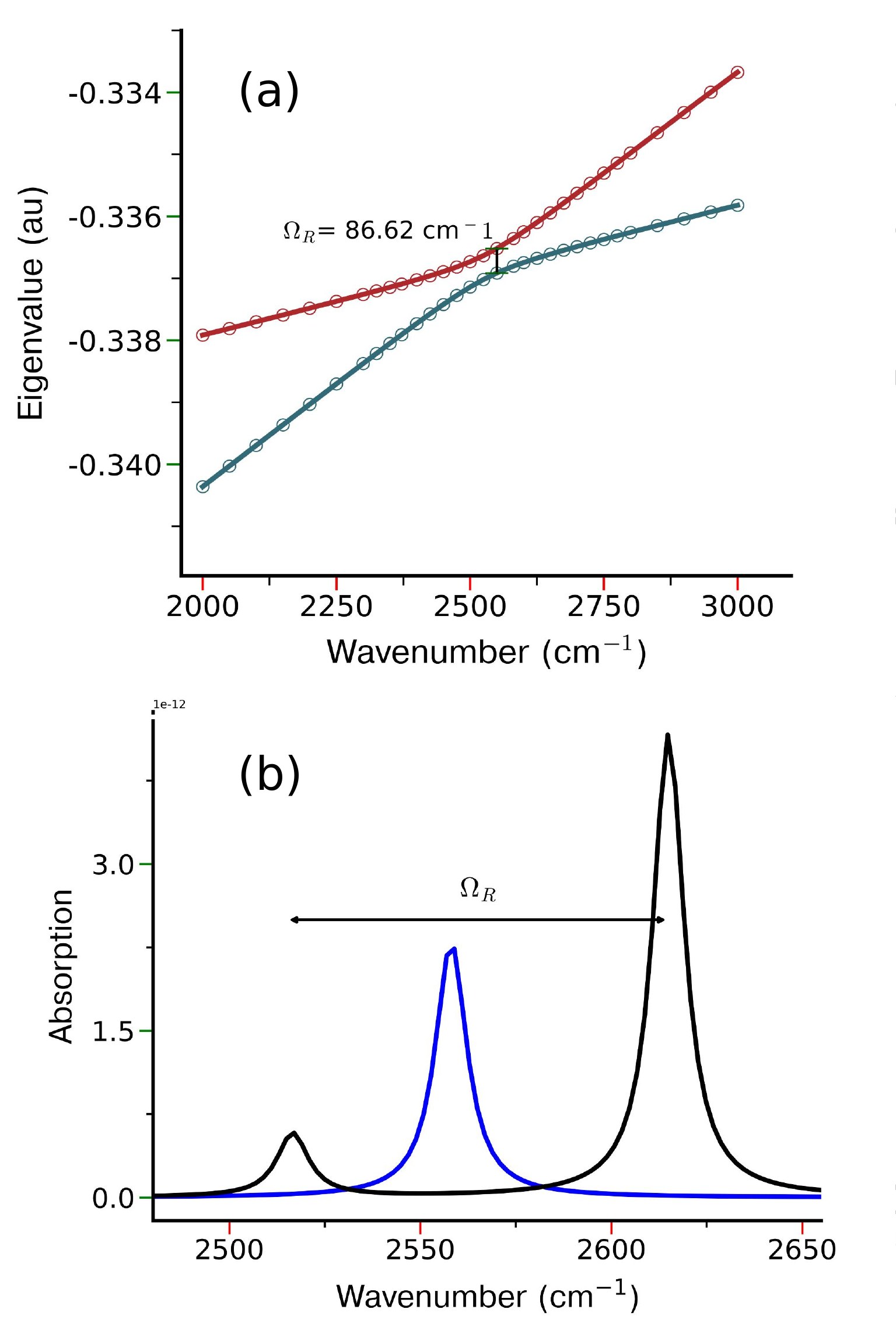}
    \caption{(a) Variation of the $2^{\mathrm {nd}}$ and $3^{\mathrm {rd}}$ eigenvalues of Eqn.~\ref{modvscham} with cavity frequency $\omega_{c}$ for parameter value $V^{\ddagger}_{2}$ exhibiting avoided crossings. (b) show the corresponding absorption spectra\cite{li2021cavity} in the presence (black) and absence (blue) of cavity-molecule coupling. The cavity-system coupling strength is fixed at $\lambda_{c} = 0.1$ au.}
    \label{fig:rabi_lamb0p1_V2}
    \end{center}
    \end{figure}

The coupling of the cavity mode to the molecular modes is via the dipole function $\mu({\mathbf r})$, and in this work, for simplicity,  we assume mainly a dependence on the reaction coordinate $x$, unless stated otherwise. In addition, we take a linearized dipole approximation $\mu(x) \approx \gamma x$, as is usually done in several such model studies on VSC. The constant $\gamma$ along with the coupling term $\lambda_{c}$ in Eq.~\ref{modvscham} determines the Rabi splitting. In Fig.~\ref{fig:rabi_lamb0p1_V2}(a), we fix $\lambda_{c}=0.1$ au, and show the variation of the lowest quantum eigenvalues of the cavity-molecule system with the cavity frequency for one of the parameter value $V_{2}^{\ddagger}$.  The avoided crossing indicates the formation of the polaritons and the corresponding absorption spectrum\cite{li2021cavity} in Fig.~\ref{fig:rabi_lamb0p1_V2}(b) for $\gamma = 1$ yields a Rabi splitting of about $\sim 100$ cm$^{-1}$, which is a typical value in the VSC regime\cite{ying2024resonance,george2023polaritonic,long2015coherent,nagarajan2021chemistry}. Therefore, in this study, we fix $\gamma=1$ within the linearized dipole approximation. It is important to note that the value $\lambda_{c} = 0.1$ in a single molecule limit is unrealistic and such values for the coupling strength are typically assumed in single molecule studies to mimic the collective strong coupling regime in VSC experiments. Alternatively, the required field strength can arise due to a large dipole or small effective mode volumes of the cavity. Thus, since $\lambda_{c} = (\epsilon_{0} V)^{-1/2}$, one estimates a cavity volume $V \sim 186~\text{\AA}^{3}$ to achieve such coupling strengths in the single molecule limit. In addition, the asymmetric  polariton bands in Fig.~\ref{fig:rabi_lamb0p1_V2}(b) arise due to the anharmonicity, asymmetric PES,  as well as the fact that the lower and upper polaritons having different transition dipole strengths.

Following earlier works\cite{fischer2022cavity,mondal2022dissociation,mondal2023phase} we choose the initial state to be a polariton wavepacket
$\Psi_{0}(\mathbf{r},q_{c};t=0) = \psi(\mathbf{r};\mathbf{r}_{0}) \phi(q_{c})\label{int_wavefunction}$ with  
\begin{equation}
\psi(\mathbf r;\mathbf r_0)
=
\left(\frac{1}{\pi \sigma_x \sigma_y}\right)^{1/2}
\exp\!\left[
-\frac{( x- x_0)^2}{2\sigma_x^2} 
-\frac{( y- y_0)^2}{2\sigma_y^2} 
\right],
\end{equation} 
being a minimum uncertainty wavepacket centered at ${\mathbf r}_{0} = (x_{0},y_{0})$ with width $\sigma_{x,y}=\left(\hbar/m\omega^{x,y}_{\mathrm R}\right)^{1/2}$ represented as a product of the displaced ground state wavefunction of the harmonized reactant well (along the reaction coordinate $x$) of frequency $\omega^{x}_{\mathrm R}$ and the ground state wavefunction along the orthogonal coordinate $y$ of frequency $\omega^{y}_{\mathrm R}$. 
The cavity state 
\begin{equation}
    \phi(q_{c}) = \left(\frac{\omega_{c}}{\pi \hbar}\right)^{1/4}\exp\left[-\frac{\omega_{c}q_{c}^{2}}{2 \hbar}\right] \label{cavityvirtual}
\end{equation}
is taken to be the ground state of the harmonic oscillator with frequency $\omega_{c}$ and unit mass. Note that for $\lambda_{c} \neq 0$, the above choice of the initial cavity state will lead to the presence of virtual photons inside the cavity. A more appropriate choice\cite{mondal2023phase} that ensures a zero photon  cavity  state is given by 
\begin{equation}
    \phi(q_{c};\lambda_{c}) = \left(\frac{\omega_{c}}{\pi \hbar}\right)^{1/4}\exp\left[-\frac{\omega_{c}(q_{c}-q_{c0})^{2}}{2 \hbar}\right] \label{cavityinit}
\end{equation}
with $q_{c0}=-\frac{\lambda_{c}}{\omega_{c}}\mu$. In the following section, we present the results for the virtual photon case. However, qualitatively similar results are obtained for the zero photon case (see Supplementary Sec. S$5$) as well. The time-evolved state $\Psi(\mathbf{r},q_{c};t)$ is obtained by numerically solving the Sch\"{o}dinger equation using the split-operator method. For classical simulations, an ensemble of initial conditions is sampled from the Wigner distribution corresponding to the initial polaritonic wavepacket, and each initial phase space point is then time-evolved by integrating Hamilton's equations of motion (see Supplementary Sec. S$7$ for details). 

\subsection{Observable of interest: branching ratio}
\label{bratio}

The key observable of interest in the context of VSC is the product branching ratio, which is as a measure of the selectivity between the two products P$_{1}$ and P$_{2}$.  In order to calculate the branching ratio one needs to define appropriate domains $D$ to demarcate reactant and product regions. We use the following definition: 
\begin{align}
    D = \begin{cases}
        \mathrm{R}, \,\, x < 0 \\
        \mathrm{P}_{1}, \,\, x > 0 \, ; \, y < y_{\mathrm{TS}_{2}}\\
        \mathrm{P}_{2}, \,\, x > 0 \, ;  \,y > y_{\mathrm{TS}_{2}}
    \end{cases}
\end{align}
where the TS$_{2}$ is located at $(x_{\mathrm{TS}_{2}},y_{\mathrm{TS}_{2}})$. Although, alternative definitions for $D$ are possible,  the choice above is made for numerical convenience and the results are qualitatively similar with different definitions for the domains.
Furthermore, given that our model has no dissipation, we also impose an energetic criterion\cite{tao2023symmetric} to mimic the reduction in any thermally induced recrossing between the two product wells. Thus, after the first crossing of TS$_{1}$, if a trajectory enters either the P$_{1}$ or P$_{2}$ domain and its potential energy falls to $\sim 3 k_{B}T$ below the TS$_{2}$ energy then the trajectory is considered to have formed that specific product.   In this work, we set $T = 300$ K, implying $3 k_{B} T \sim 2.85 \times 10^{-3}$ au, and note that the main results of this work are fairly insensitive to the choice of this cutoff (see Supplementary Fig. S$7$).

    \begin{figure*}[htbp]
    \centering  
    \includegraphics[width=1.0\linewidth]{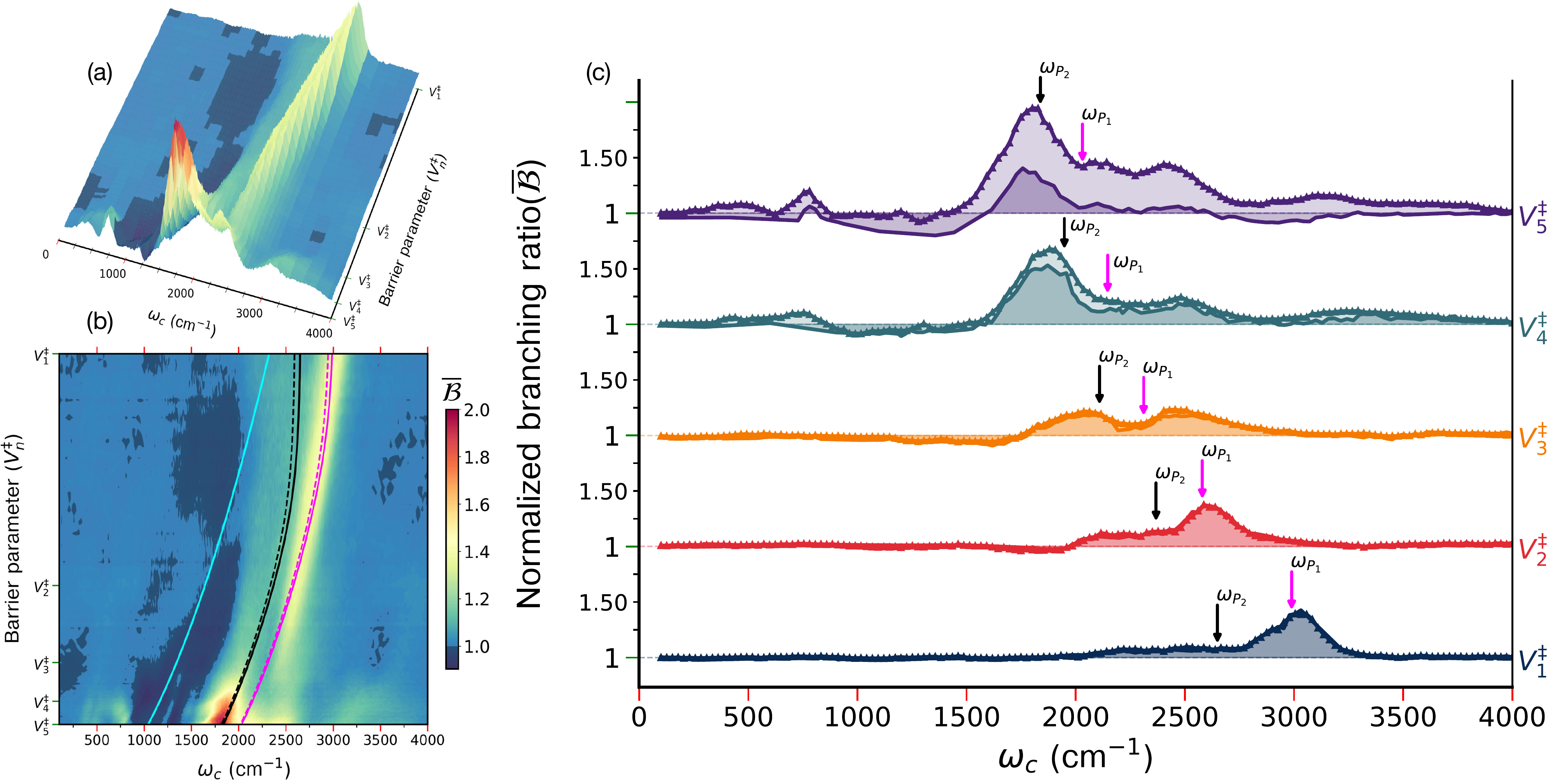}
    \caption{(a) and (b) Two different representations of the classical normalized branching ratio Eq.~\ref{normbranchrat} as a function of the cavity frequency $\omega_c$ and barrier parameter $V^{\ddagger}_n$. The landscape is computed on a $311 \times 100$ grid (see supplementary for details on the grid setup) for an initial state with energy $\langle E_{S}\rangle \approx 0.01$ au with the cavity-system coupling strength $\lambda_{c} = 0.1$ au and asymmetry parameter $\alpha = 0.1$. In (b) the varying harmonic (solid lines) and fundamental (dashed lines) frequencies  of the reactant well $\omega_{\mathrm{R}}^{x}$ (cyan), the deeper product well $\omega_{\mathrm{P}_{1}}^{x}$ (magenta), and the product well $\omega_{\mathrm{P}_{2}}^{x}$ (black) are superimposed.  (c) Classical (lines with symbols) and quantum (solid lines) branching ratios for specific $V^{\ddagger}_{n}$ values. The $x$-mode harmonic frequencies for the two product wells are indicated with arrows.}
    \label{fig:xcoupledBR}
    \end{figure*} 
    
With the definitions above, we calculate the classical time-dependent domain probability as 
    \begin{equation*}
        \mathrm{Prob}^{CM}_{D}(t) = \frac{\mathcal{N}_{D}(t)}{\mathcal{N}_{tot}} \label{pd_cm}
    \end{equation*}
    where $\mathcal{N}_{D}(t)$ is the number of trajectories in the domain $D$ and  $\mathcal{N}_{tot}$ is the total number of trajectories. Quantum mechanically, the corresponding domain probability is given by 
    \begin{equation}
      \mathrm{Prob}^{QM}_{D}(t)=\iint_D|\Psi(\mathbf{r},q_{c};t)|^{2} \,  d\mathbf{r} \,dq_{c}\  \label{pd_qm}
    \end{equation}
    The respective time-averaged domain probabilities are calculated as follows
    \begin{equation}
        \overline{\cal P}_{D} = \frac{1}{T_{\mathrm tot}}\int_{0}^{T_{\mathrm tot}} \mathrm{Prob}_{D}(t) \, dt   \label{pd_cm} 
    \end{equation}
    where $T_{\mathrm tot}$ is the total propagation time.  In the results presented below, we take $T_{\mathrm tot} = 1$ ps and note that qualitatively similar results are obtained for larger $T_{\mathrm tot}$ as well (see Supplementary Sec. S$2$). The time-averaged domain probabilities allow us to define the branching ratio (selectivity)  
      \begin{equation}
        {\mathcal{B}}(\bar{V}^{\ddagger},\lambda_{c},\omega_{c}) = \frac{\overline{\cal P}_{\mathrm{P}_{1}}}{\overline{\cal P}_{\mathrm{P}_{2}}}  \label{br}
    \end{equation}
      Since the asymmetry parameter is fixed at $\alpha = 0.1$, $\mathcal{B}$ can be different from unity even in the absence of coupling ($\lambda_{c} = 0$, see Supplementary Fig. S$5$) to the cavity. Therefore, we define the normalized branching ratio
    \begin{equation}
        \overline{\mathcal{B}}(\bar{V}^{\ddagger},\lambda_{c},\omega_{c}) \equiv \frac{\mathcal{B}(\bar{V}^{\ddagger},\lambda_{c},\omega_{c})}{\mathcal{B}(\bar{V}^{\ddagger},0,\omega_{c})}
        \label{normbranchrat}
    \end{equation}
     as a measure of the extent to which VSC modulates the branching ratio.

\section{Results and discussion}
\label{results}

In what follows, we choose the center of the matter wavepacket  $(x_{0},y_{0}) = (-0.85,0.016)$ au, corresponding to the average system energy $\langle E_{S}\rangle = \langle \psi|H_{M}|\psi\rangle \approx 0.01$ au. This energy is above that of the TS$_{1}$ saddle ($E_{\mathrm{TS}_{1}} = 0$) and chosen in order to facilitate a classical-quantum comparison of the VSC effect. In addition, typical PTSB reactions involve "hot" reactants crossing the TS$_{1}$ towards the products and our choice of energy is corresponding to such a situation. Although, for the specific choice of the initial wavepacket, $\langle E_{S}\rangle$ does depend on the parameter $V^{\ddagger}_{n}$, the variations are small over the range of interest (see Supplementary Fig. S$4$).  Note that the main features of our results remain invariant for other choices of $\langle E_{S}\rangle$ values (Supplementary Sec. S$3$).

    \begin{figure*}[htbp]
    \centering
    \includegraphics[width=17.1 cm]{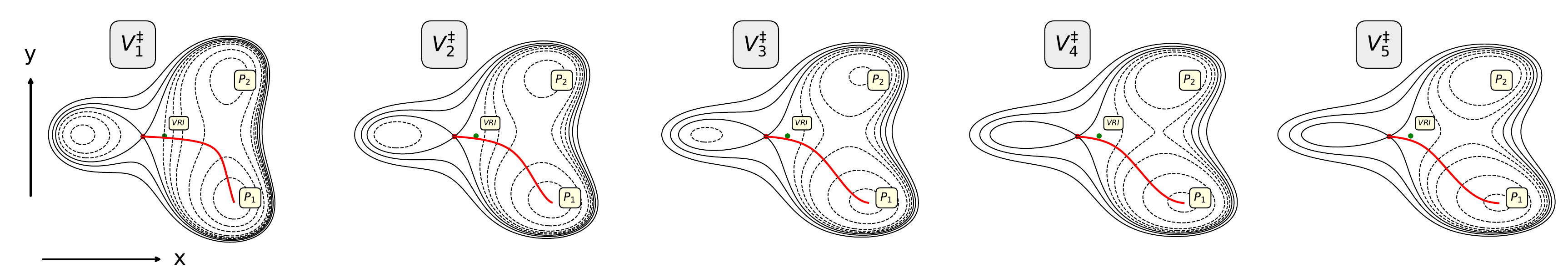}
    \caption{Intrinsic reaction coordinate (IRC, red curve) projected onto the two-dimensional PES for varying barrier parameter $\overline{V}^\ddagger = V_{n}^{\ddagger}/x_{s}^{4} \equiv V^{\ddagger}/n x_{s}^{4}$ with $n=1,\ldots,5$.}
    \label{fig:irc_varyVdag}
    \end{figure*}

\subsection{Establishing the VSC effect}

    The central results of our computations are summarized in Fig.~\ref{fig:xcoupledBR} wherein the classical branching ratio $\overline{\mathcal{B}}(\bar{V}^{\ddagger},\lambda_{c},\omega_{c})$ is shown as a function of the cavity frequency $\omega_{c}$ and the parameter $V^{\ddagger}_{n}$ which controls the depths and the fundamental frequencies of the reactant and product wells. Essentially, Figs.~\ref{fig:xcoupledBR}(a) and (b) show the ``VSC landscape" for our model system and one can clearly observe specific regions on this landscape exhibiting significant variations in the branching ratio.  Note that with $\lambda_{c} = 0.1$ we have the scaled coupling strength parameter $\eta \equiv \Omega_{R}/2\omega_{c}$ being in the strong coupled regime for most of the  results shown below. Nevertheless, variations in $\eta$ values are expected across the range of cavity frequencies of interest. Supplementary Sec. S$3.3$ shows that the results for fixed $\eta$ values are essentially the same. Three key observations can be made: 
    \begin{itemize}
        \item A ridge of enhanced $\overline{\mathcal{B}}$ can be seen starting from $V^{\ddagger}_{1}$ at around $\omega_{c} \sim 3000$ cm$^{-1}$ and persisting up until nearly $V^{\ddagger}_{3}$ with $\omega_{c} \sim 2500$ cm$^{-1}$.
        \item Beyond $V^{\ddagger}_{3}$, as the barrier parameter is decreased, the ridge bifurcates with a prominent, and maximal for this study, enhancement in the branching ratio around $\omega_{c} \sim 1800$ cm$^{-1}$ for $V^{\ddagger}_{5}$.
        \item  The comparison between classical and quantum $\overline{\mathcal{B}}$ for select $V^{\ddagger}_{n}$ shown in Fig.~\ref{fig:xcoupledBR}(c) indicates that the initial ridge structure is reproduced quantum mechanically as well. Importantly, despite quantitative differences, the cavity frequencies at which significant modulation of $\overline{\mathcal{B}}$ occurs remain nearly invariant in the quantum and classical cases.
    \end{itemize}

     Are the  modulations of selectivity noted above due to VSC? The answer is in the affirmative, and this can be seen from Fig.~\ref{fig:xcoupledBR}(b) where we have superimposed the $x$-mode harmonic frequencies corresponding to the reactant and the two products onto the VSC landscape. It is immediately clear that the initial ridge structure corresponds to the resonance $\omega_{c} \sim \omega_{\mathrm{P}_{1}}^{x}$ between the cavity frequency and the harmonic frequency of the $\mathrm{P}_{1}$ product well. Significantly, the ridge structure nearly follows the variation of $\omega_{\mathrm{P}_{1}}^{x}$ with $V^{\ddagger}_{n}$ seen in Fig.~\ref{fig:harmfreq_variationwithVdag}, thus implying the robustness of the resonance effect. Interestingly, the bifurcation of the ridge leading to a shift of the maximum in $\overline{\mathcal{B}}$ starting from $V^{\ddagger}_{4}$ corresponds to the resonance $\omega_{c} \sim \omega_{\mathrm{P}_{2}}^{x}$ involving the cavity mode and the harmonic frequency of the shallower $\mathrm{P}_{2}$ product well. This observation is surprising, and perhaps counterintuitive as well, since VSC of the cavity mode with the shallower well (``minor product") results in an increased selectivity of the deeper well (``major product"). Thus, depending on the relative depths of the product wells, the VSC effect exhibits a switch in the resonant frequency.  At this stage it is worth mentioning that the observed resonant enhancements of $\overline{\mathcal{B}}$ are robust under variations of various system parameters (see Supplementary sec. S$2$ and S$3$) and the off-cavity branching ratio varies very little over the entire range of $V^{\ddagger}_{n}$ (see Supplementary Fig. S$5$). 
    
     In order to examine the VSC effect in more detail, Fig.~\ref{fig:xcoupledBR}(c) shows slices of the landscape at specific $V^{\ddagger}_{n}$ values. For $V^{\ddagger}_1$, the normalized branching ratio $\overline{\cal B}$ shows a peak around $\omega_{c} \sim \omega_{\mathrm{P}_{1}}^{x}$ and the peak persists at $V^{\ddagger}_2$, but with slightly reduced intensity. For $V^{\ddagger}_3$, the picture becomes more complex with $\overline{\cal B}$ exhibiting two peaks corresponding to nearly the same enhancement. The second new peak nearly corresponds to $\omega_{c} \sim \omega_{\mathrm{P}_{2}}^{x}$ resonance whereas the first peak is blue-shifted from the  $\omega_{c} \sim \omega_{\mathrm{P}_{1}}^{x}$ resonance, indicating competing resonant pathways. At $V^{\ddagger}_4$ the maximal enhancement of $\overline{\mathcal{B}}$ appears around the resonance $\omega_{c} \sim \omega_{\mathrm{P}_{2}}^{x}$ and this trend is amplified at $V^{\ddagger}_5$, where the branching ratio is nearly twice that of the off-cavity case. In addition, for both the $V^{\ddagger}_{4}$ and $V^{\ddagger}_{5}$ cases the enhancement in selectivity at $\omega_{c} \sim \omega_{\mathrm{P}_{1}}^{x}$ appears to be part of a rather broad shoulder-like feature. In addition, as seen from Fig.~\ref{fig:xcoupledBR}(b) as well as Fig.~\ref{fig:harmfreq_variationwithVdag}, since the harmonic frequencies corresponding to the two product wells are fairly close to their fundamental frequencies, the observed resonance shift is not due to anharmonic effects. 
     Importantly, at all values of $\overline{V}^{\ddagger}$ the system continues to prefer P$_1$; what changes is the cavity frequency at which this preference is most strongly enhanced.

     Another key point to be noted at this stage is that the branching ratios in Fig.~\ref{fig:xcoupledBR}(c) exhibit good classical-quantum correspondence in terms of the resonance frequencies as well as the shift of the dominant enhancement frequency. In fact, even the domain probabilities (see Supplementary Fig. S$14$) show good classical-quantum correspondence.
     Consequently, this indicates that the primary mechanism governing the cavity-modified selectivity is largely classically in origin and arises from the dynamical steering in the post transition state region. Nevertheless, quantitative differences between classical and quantum $\overline{\cal B}$ can be seen for $V^{\ddagger}_{4}$ and $V^{\ddagger}_{5}$ with the latter exhibiting a reduced enhancement for $\omega_{c} \sim \omega_{\mathrm{P}_{2}}^{x}$ when compared to the classical result.  A possible explanation for this has to do with quantum tunneling between the product wells and the resulting partial delocalization of the wavepacket. However, in this work we do not further analyze such effects.

    \subsection{Mechanism for the enhancement of selectivity}
    \label{howandwhy}

    A possible explanation for the results shown in Fig.~\ref{fig:xcoupledBR} involves a cavity coupling mediated changes in the nature of the intrinsic reaction coordinate (IRC).  Indeed, as shown in Fig.~\ref{fig:irc_varyVdag}, varying $V^{\ddagger}_{n}$ results in the IRC moving away from TS$_2$ and almost exclusively leading to the product $\mathrm{P}_{1}$. If this assumption holds, then one should observe a monotonic increase in the selectivity $\overline{\mathcal {B}}$ at the cavity-molecule resonance $\omega_{c} \sim \omega_{\mathrm{P}_{1}}^{x}$. However, the results shown in Fig.~\ref{fig:xcoupledBR}(c) contradict this expectation. In particular, apart from the reduction in $\overline{\mathcal {B}}$ for $V^{\ddagger}_{3}$, the observation of a different resonance $\omega_{c} \sim \omega_{\mathrm{P}_{2}}^{x}$ leading to the maximal enhancement in the selectivity of $\mathrm{P}_{1}$ for $V^{\ddagger}_{4}$ and $V^{\ddagger}_{5}$ cannot be explained by arguments based solely on the change in the IRC.  Thus, the resonant modulations in the selectivity of the coupled cavity-system case must be due to dynamical effects. 
    
    An important clue emerges from Fig.~\ref{fig:xcoupledP1P2} wherein the time averaged product well domain probabilities are shown (see Supplementary Fig. S$13$ for more details on the domain and recrossing probabilities). These results should be compared to the branching ratios shown in Fig.~\ref{fig:xcoupledBR}(c) since it is the ratio of the domain probabilities that yield the branching ratios. For cavity frequencies away from the resonant frequencies, Fig.~\ref{fig:xcoupledP1P2} shows that the $\overline{\cal P}_{D}$ for both the product wells is influenced by the cavity but varies to a similar extent. Thus, for such cavity frequencies, the branching ratio in Fig.~\ref{fig:xcoupledBR}(c) remains close to unity. However, for $\omega_{c}$ near the TS$_{2}$ transverse frequency $\omega^{x}_{\mathrm{TS}_{2}}$, the $\mathrm{P}_{2}$ well domain probabilities begin to decrease while the $\mathrm{P}_{1}$ well domain probabilities keep increasing until $\omega_{c} \sim \omega^{x}_{P}$. Consequently, the substantial difference in the domain probabilities of the two wells at $\omega_{c} \sim \omega^{x}_{P}$ gives rise to the resonant enhancements seen in Fig.~\ref{fig:xcoupledBR}(c). Given that earlier studies\cite{sun2022suppression,ke2025non,mondal2023phase} have emphasized the key role of the cavity in post transition state cooling into the product well, the results in Fig.~\ref{fig:xcoupledP1P2} suggest that analyzing the resonant cavity-system energy exchange is necessary to rationalize the observed enhancement in selectivity. 
    
     \begin{figure}[t]
    \centering  
    \includegraphics[width=8.6cm,height=8.4cm]{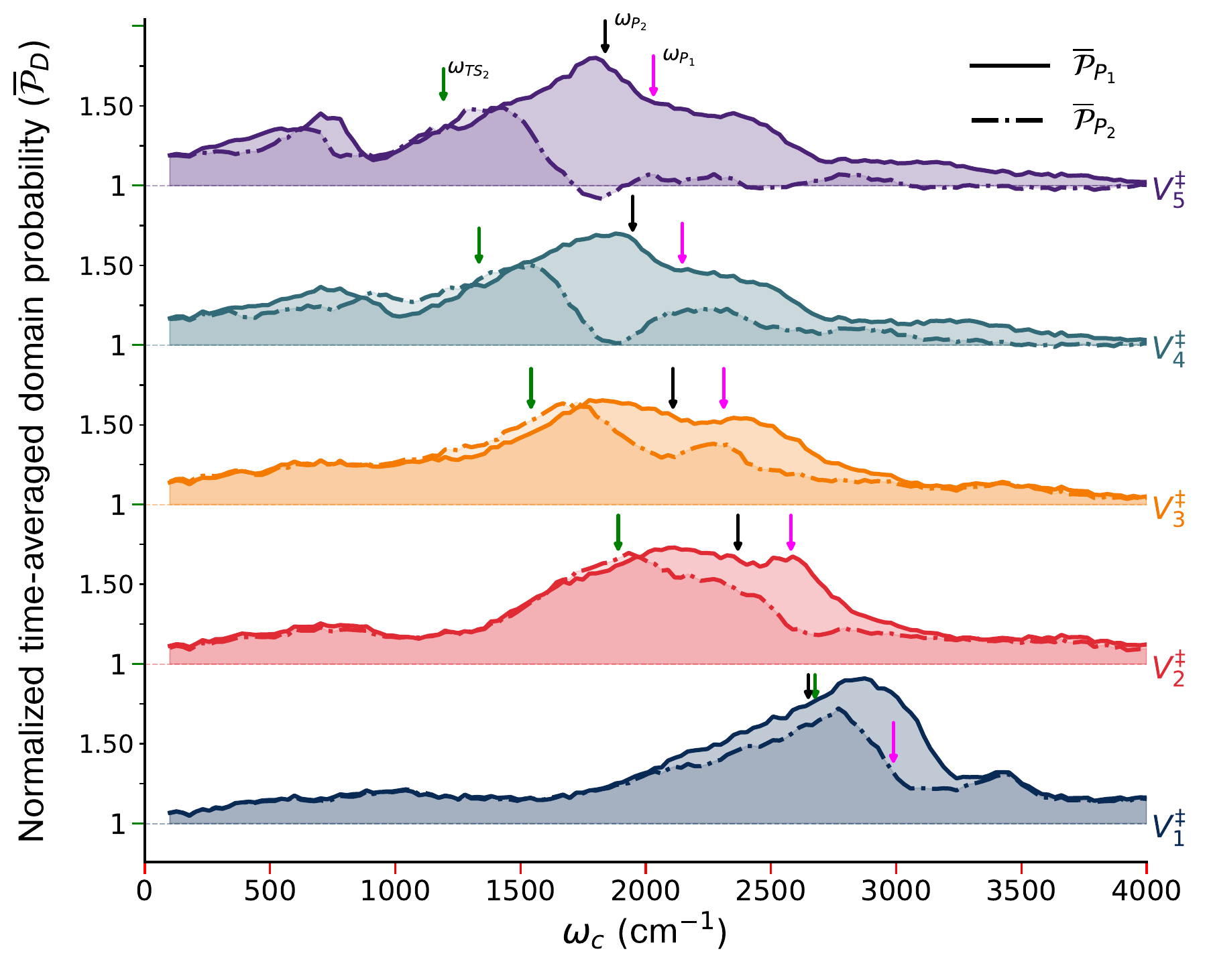}
    \caption{Classical normalized time averaged domain probabilities Eq.~\ref{pd_cm} corresponding to the branching ratios shown in \ref{fig:xcoupledBR}(c). Apart from the $x$-mode frequencies of the product wells, the transverse mode frequency at TS$_{2}$ is also indicated (green arrow).}
    \label{fig:xcoupledP1P2}
    \end{figure}

    \begin{figure*}[htbp]
    \centering
    \includegraphics[width=13.4 cm]{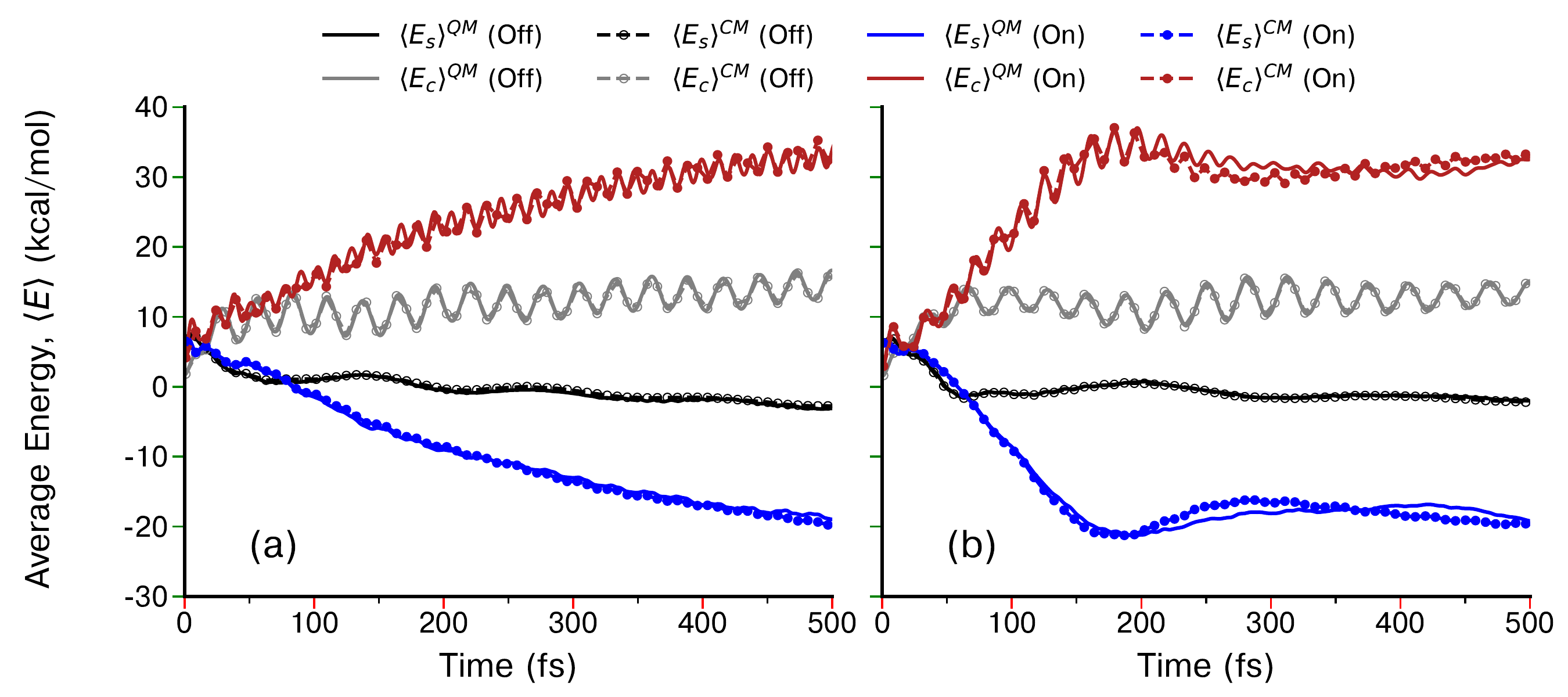}
    \caption{Average energies (indicated in the legend) as a function of time for  (a) $V^{\ddagger}_{2}$ with $\omega_{c}=1200$ cm$^{-1}$ (off resonance), $\omega_{c}=\omega_{P_{1}}^{x} = 2586$ cm$^{-1}$ (on resonance) and (b) $V^{\ddagger}_{4}$ with $\omega_{c}=1071$ cm$^{-1}$ (off resonance),  $\omega_{c}=\omega_{P_{2}}^{x} = 1871$ cm$^{-1}$ (on resonance). Solid lines  and dotted lines with markers show the quantum and classical results respectively.}
    \label{fig:avgE_QMCM_V2V4}
    \end{figure*}

     \begin{figure*}[htbp]
    \centering
    \includegraphics[width=15.4 cm]{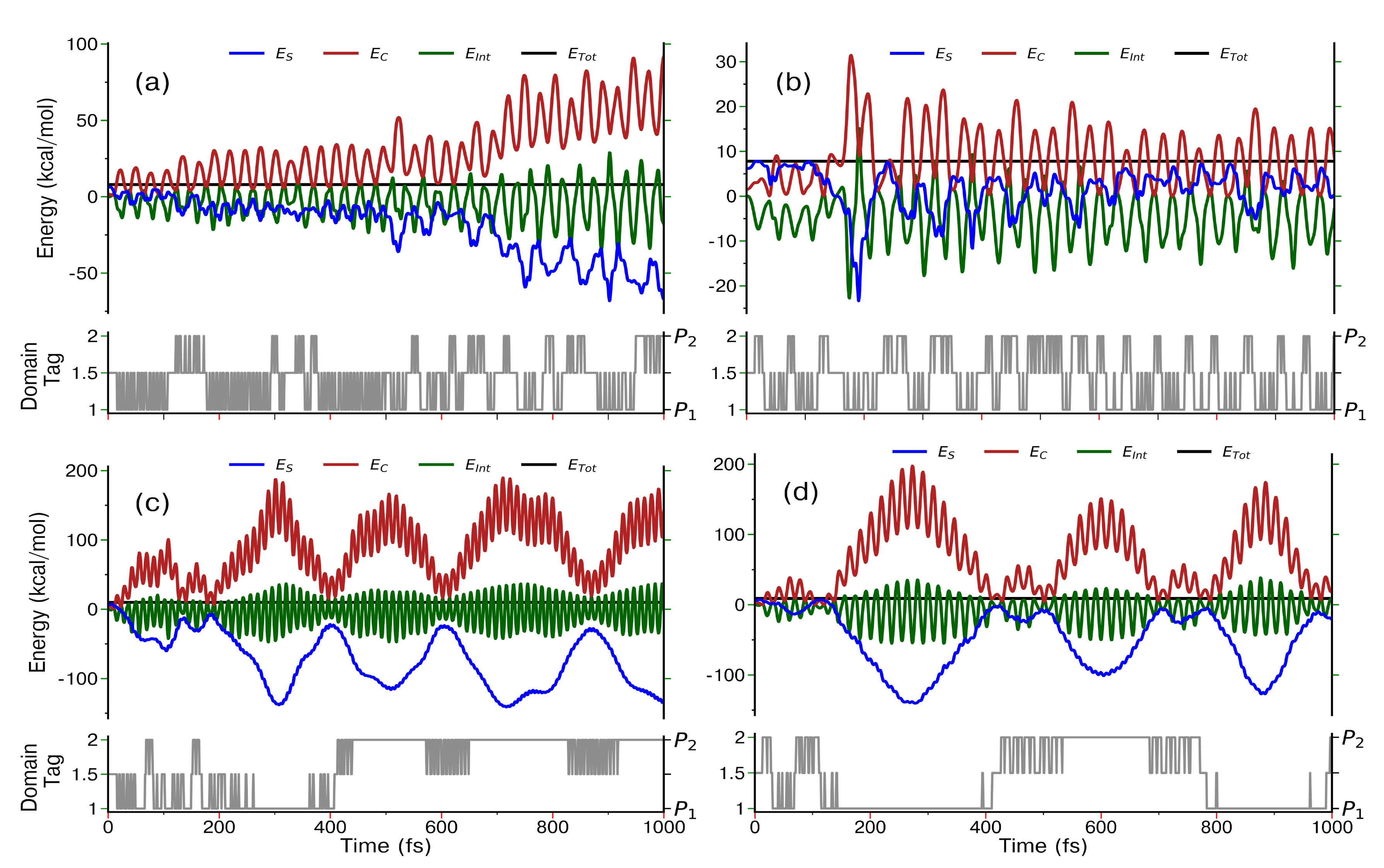}
    \caption{Energy variations for sample classical trajectories as a function of time.  The system, cavity, and interaction energies are shown in blue, red, and green respectively. (a) $V^{\ddagger}_{2}$ with cavity frequency $\omega_{c}=1200$ cm$^{-1}$ (off resonance), (c) $V^{\ddagger}_{2}$ with $\omega_{c}=\omega_{P_{1}}^{x} = 2586$ cm$^{-1}$ (on resonance), (b) $V^{\ddagger}_{4}$ with $\omega_{c}=1071$ cm$^{-1}$ (off resonance),  and (d) $V^{\ddagger}_{4}$ with $\omega_{c}=\omega_{P_{2}}^{x} = 1871$ cm$^{-1}$ (on resonance). For each trajectory we associate (lower panels)
     domain tag values $1,2$, and $0$  whenever it is in the $P_{1}, P_{2}$, and $R$ wells respectively. The domain tag value $1.5$ corresponds to the case when the trajectory is outside all three wells.
    All computations are shown for cavity-system coupling strength $\lambda_{c} = 0.1$ au and $\alpha=0.1$}
    \label{fig:singletraj_V2V4}
    \end{figure*}
    
    \subsubsection{Coherent cavity-system energy exchange}\label{coherenteexchange}
    
    In order to examine the cavity-system energy exchange dynamics, we begin by showing in Fig.~\ref{fig:avgE_QMCM_V2V4}(a) and (b) the average system and cavity energies (see Supplementary Sec. S$6$ for definitions) in the off and on resonance scenario for $V^{\ddagger}_{2}$ and $V^{\ddagger}_{4}$ cases respectively. Note that both classical and quantum results are compared with appropriate initial states that were utilized to obtain the branching ratios shown in Fig.~\ref{fig:xcoupledBR}. Apart from the striking classical-quantum correspondence, one can observe a clear difference in the average energies when the cavity is on and off resonant with the product well $x$-mode frequencies. While there is some cavity-system energy exchange in the off-resonant cases, the on-resonance cases show a substantial cooling of the system over timescales of a few hundred femtoseconds. This cooling of the system clearly correlates with the cavity gaining energy and hence establishes the active cavity-system energy exchange for $\omega_{c} \sim \omega^{x}_{\mathrm{P}_{1}}$ and $\omega_{c} \sim \omega^{x}_{\mathrm{P}_{2}}$ cases. Note that for the choice of the off-resonant frequencies in Fig.~\ref{fig:avgE_QMCM_V2V4}(a) and (b), one observes, despite the influence of the cavity on the domain probabilities in Fig.~\ref{fig:xcoupledP1P2}, a near unit branching ratio. 
    
    For gaining further insights into the on-resonance cooling of the system seen in Fig.~\ref{fig:avgE_QMCM_V2V4}(a) and (b), we analyze the detailed classical dynamics of the model system.  An ensemble of $20000$ initial trajectories is sampled at the first transition state TS$_{1}$ with the condition $p_{x} > 0$ for a fixed system energy of $E_{s}=0.01$ au and the cavity mode being in its ground state. Thus, in terms of the cavity action-angle variables, the cavity action is fixed at $J_{c}=0.5$, while the angle variable $\theta_{c}$ is randomly chosen from the interval $\left[0,2\pi\right]$. We then propagate Hamilton’s equations of motion, ensuring proper convergence of the results.

    In Fig.~\ref{fig:singletraj_V2V4}(a)-(d), we show the time evolution of the various energies for representative example trajectories in on and off resonance cases.  Note that similar qualitative behavior is seen for several other trajectories as well. To track the time evolution in terms of the various domains that are visited, we assign a ``domain tag" which is assigned values of $1,2$, and $0$ when the trajectory enters the P$_{1}$, P$_{2}$, and the $\mathrm{R}$  wells, respectively. Further, a value of $1.5$ is assigned when the trajectory is not in any of the three wells, but in some intermediate regions. Such an analysis is useful in providing information, not readily available from the average energy results shown in Fig.~\ref{fig:avgE_QMCM_V2V4}, about the regions where dominant cooling of the system is observed. The off-resonance cases for $V^{\ddagger}_{2}$ and $V^{\ddagger}_{4}$ are shown in Figs.~\ref{fig:singletraj_V2V4}(a) and (b), respectively, along with the domain tag of the trajectory. In these cases, it is clear that the energy exchanged between the system and the cavity mode is relatively small and incoherent. The domain tags show frequent interwell transitions between the two product wells and no visits to the reactant well over the timescale of interest.  Note that recrossing into the reactant regions can be observed in other trajectory examples. However, as seen in Supplementary Fig. S$13$(c), in the on-resonance scenario the recrossing probability is reduced on average, consistent with the longer dwell times in the product wells due to extensive cooling. In contrast, under on-resonance conditions shown in Figs.~\ref{fig:singletraj_V2V4}(c) and (d), one can clearly see significant and coherent cavity-system energy transfer over a timescale of $\sim 200$ fs. In both on and off resonant cases the fast oscillations can be associated with the cavity frequency. On the other hand, given that in the $V_{2}^{\ddagger}$ case the time scale associated with the Rabi splitting of $\sim 86$ cm$^{-1}$ (cf. Fig.~\ref{fig:rabi_lamb0p1_V2}(a)) is around $\sim 400$ fs,  the coherent cavity-molecule energy exchanges are occurring on timescales comparable to the Rabi timescale. Similar observations hold for the $V_{4}^{\ddagger}$ case as well. However, in such  anharmonic  systems the resonant energy exchange process is more complicated. For instance, in the on-resonance cases one can clearly see that the periods of the coherent energy oscillations does change with time, hinting at the involvement of more than two states. More importantly, when the trajectory enters either of the product wells, it remains trapped for an extended duration in that well, as indicated by the domain tag, during which a substantial amount of energy is transferred from the system to the cavity mode, leading to cooling. Subsequently, as has been observed earlier\cite{GarnerJCP2025}, the cavity returns the energy back to the system as the trajectory exits out of that specific product well.  These repeated, synchronized energy transfer events indicate a coherent exchange between the reaction coordinate and the cavity mode during each trapping episode. An interesting consequence of our computations is that, irrespective of the well which is involved in VSC, substantial cooling occurs in both the product wells.

     \begin{figure*}[htbp]
    \centering  
    \includegraphics[width=15.4cm]{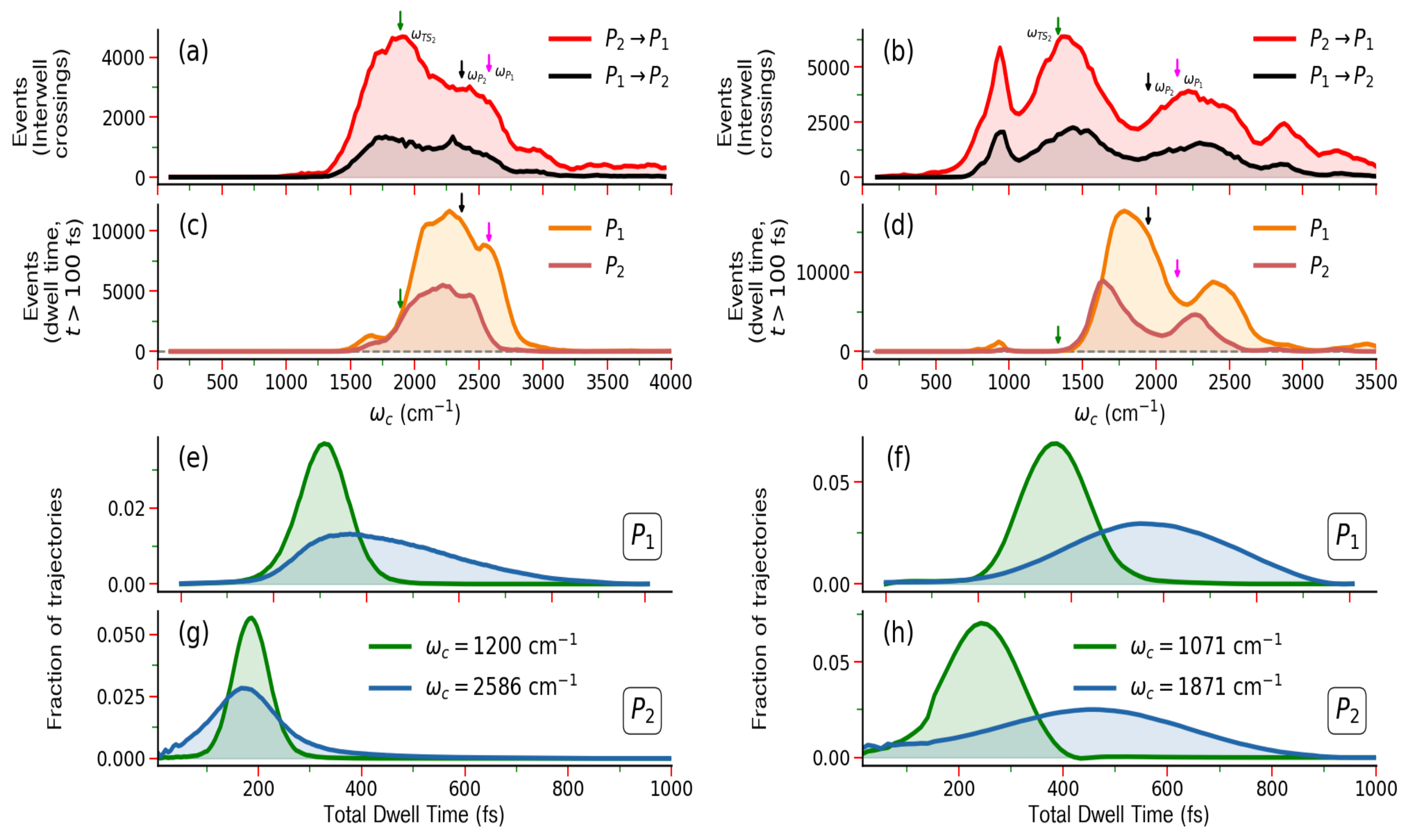}
    \caption{(a) and (c) Number of trajectories undergoing P$_{1} \leftrightarrow$ P$_{2}$ inter-well transitions and exhibiting dwell times in the product wells for at least $100$ fs respectively as a function of the cavity frequency for $V^{\ddagger}_{2}$. (b) and (d) show the corresponding results for  $V^{\ddagger}_{4}$ case. (e) and (g) show the product well dwell time distributions for $V^{\ddagger}_{2}$ in the resonant (blue) and off-resonant (green) cases. The corresponding results for $V^{\ddagger}_{4}$ are shown in (f) and (h). Note that for (e)-(h) an ensemble size of $10^{5}$ trajectories is used.}
    \label{fig:interwell_coolingwithP1P2_V2V4}
    \end{figure*}
    
    \subsubsection{Insights into the VSC frequency switching}
     \label{dwelltime}

    A key observation from Fig.~\ref{fig:xcoupledBR}(c) was the fact that the branching ratio peak switched from $\omega_{c} \sim \omega^{x}_{\mathrm{P}_{1}}$ to $\omega_{c} \sim \omega^{x}_{\mathrm{P}_{2}}$ as the barrier parameter is tuned from $V^{\ddagger}_{1}$ to $V^{\ddagger}_{5}$. This, as pointed out above, approximately correlates with the observations in Fig.~\ref{fig:xcoupledP1P2} wherein, in the vicinity of the TS$_{2}$ transverse frequency, the domain probability of the shallower P$_{2}$ well decreases while the deeper well domain probability continues to increase. Given that both wells undergo cooling, and hence trap population, this suggests that interwell P$_{1}\leftrightarrow$P$_{2}$ transitions are playing a key role with changing $V^{\ddagger}_{n}$.

    In order to highlight the role of the interwell transitions, in Fig.~\ref{fig:interwell_coolingwithP1P2_V2V4}(a) and (b) we show the statistics of the P$_{1}\leftrightarrow$P$_{2}$ transitions for $V^{\ddagger}_{2}$ and $V^{\ddagger}_{4}$ cases respectively as a function of the cavity frequency. The computations are done as follows: once a trajectory enters one of the wells, an interwell crossing event is counted if the trajectory transitions to the other well and stays trapped for $\sim 20$ fs, which is the timescale associated with the TS$_{2}$ barrier curvature. It is immediately clear that for $V^{\ddagger}_{4}$ far more interwell transitions happen and, interestingly,  even the peak in the interwell transitions around $\omega_{c} \sim 900$ cm$^{-1}$ appears to correlate with the small enhancement of the branching ratio seen in Fig.~\ref{fig:xcoupledBR}(c). The reason for the increased interwell transitions is due to the change in the shape of the PES -- upon changing the barrier parameter from $V^{\ddagger}_{2}$ to $V^{\ddagger}_{4}$ the transition state TS$_{2}$ and reactant well $R$  come energetically close (see Supplementary Fig. S$2$ bottom row and S$3$) and, as seen from Fig.~\ref{fig:harmfreq_variationwithVdag}, the transverse frequency $\omega^{x}_{\mathrm{TS}_{2}}$ decreases.  This results in a relatively flat region of the PES after the VRI point and, given that cooling can occur in both wells, the increased  P$_{2} \rightarrow$P$_{1}$ transitions can lead to long time trapping in the P$_{1}$ well when $\omega_{c} \sim \omega^{x}_{\mathrm{P}_{1}}$ or $\omega^{x}_{\mathrm{P}_{2}}$. 
    This can be seen from  Fig.~\ref{fig:interwell_coolingwithP1P2_V2V4}(c) and (d) which show the fraction of trajectories that get trapped in the product wells for $100$ fs or more. Note that, by construction, the interwell transitions capture the short timescale dynamics, whereas the results in Fig.~\ref{fig:interwell_coolingwithP1P2_V2V4}(c) and (d) account for the long time dynamics due to the VSC effect.

    Insights into the frequency switching come from a closer inspection of Figs.~\ref{fig:interwell_coolingwithP1P2_V2V4}(c)-(h), where key differences between the $V^{\ddagger}_{2}$ and $V^{\ddagger}_{4}$ cases emerge. In the case of $V^{\ddagger}_{2}$ shown in Fig.~\ref{fig:interwell_coolingwithP1P2_V2V4}(c) the fraction of long time trapped trajectories peaks around $\omega_{c} \sim \omega^{x}_{\mathrm{P}_{2}}$ for both product wells. However, since P$_{1}$ is deeper than the P$_{2}$ well, the fraction is considerably larger for the deeper well. Around the same cavity frequency the P$_{1} \rightarrow$P$_{2}$ transitions decrease precipitously. At $\omega_{c} \sim \omega^{x}_{\mathrm{P}_{1}}$ resonance, the trapping or dwell time distributions shown in Fig.~\ref{fig:interwell_coolingwithP1P2_V2V4}(e) and (g) indicate that the P$_{1}$ well traps population more effectively when compared to the P$_{2}$ well. Consequently, the combination of reduced interwell transitions at this frequency and the efficient cooling in the P$_{1}$ well result in the branching ratio peak at $\omega_{c} \sim \omega^{x}_{\mathrm{P}_{1}}$ in Fig.~\ref{fig:xcoupledBR}(c). In contrast, for the $V^{\ddagger}_{4}$ case Fig.~\ref{fig:interwell_coolingwithP1P2_V2V4}(d) exhibits a peak in the long time trapped fraction around the resonance $\omega_{c} \sim \omega^{x}_{\mathrm{P}_{2}}$ whereas the corresponding population in the P$_{2}$ well is comparatively much smaller. Interestingly, Figs.~\ref{fig:interwell_coolingwithP1P2_V2V4}(g) and (h) indicate that in this instance the P$_{2}$ well can also trap population effectively at resonance. Nevertheless, the shallowness of the P$_{2}$ well leads to increased P$_{2} \rightarrow$P$_{1}$ transitions around $\omega_{c} \sim \omega^{x}_{\mathrm{P}_{2}}$, as evident from Fig.~\ref{fig:interwell_coolingwithP1P2_V2V4}(b) and hence a much more rapid loss in the fraction of long time trapped trajectories in the shallow well. The net effect can be seen in Fig.~\ref{fig:xcoupledBR}(c) as a peak around $\omega_{c} \sim \omega^{x}_{\mathrm{P}_{2}}$ in the branching ratio.

\section{Conclusion}
\label{conclusion}

The potential for VSC to modulate reactions in a mode-specific manner is crucially dependent on our understanding of the cavity-induced mechanistic changes in the reaction dynamics. A prime aspect in this context depends on our ability to correlate possible effects of VSC  with the complex features that are typically found on multidimensioanl PES. In this work, by utilizing detailed classical and quantum dynamics on a model system exhibiting post–transition state bifurcation (PTSB), we have established that VSC can influence the selectivity between major and minor products of a chemical reaction. Moreover, by varying a single parameter that controls the shape of the PES, we have illustrated the robustness of our observations and that the cavity with an appropriately tuned frequency effectively acts as an engineered vibrational bath that facilitates ``cooling down” of the system. Note that the cooling of the system under resonant VSC has been invoked before\cite{sun2022suppression,ke2025non,GarnerJCP2025} with emphasis on the dynamical nature of the rate modulations. A new aspect of our work is that the cooling occurs in both the product wells even if the cavity frequency is resonant with only one of the product well's fundamental transition.  Note that the parameter space for our model in Eq.~\ref{modelPES} is rather large and further studies are needed to see if there are regimes wherein the selectivity is maximally sensitive to VSC with the reactant well transition. Combined with recent observations\cite{CarpenterJPCA2022} by Carpenter which underscores the role of interwell transitions in PTSB systems with and without baths, the VSC induced cooling effect observed here may lead to crucial insights for developing approximate predictive models for product selectivity. 

Apart from a good classical-quantum correspondence for the branching ratios, our study also underscores the complexity of the VSC phenomenon. To begin with, the cavity frequency at which a maximal VSC response is expected depends on the shape of the PES. This seems to be related to the fact that both the product wells can trap population under a single resonance condition. Although the mechanism for this is not entirely clear at present,  we suspect large scale changes to the classical phase space structures and the importance of resonance between the nonlinear mode frequencies of the system and the cavity frequency. Further detailed studies are underway to confirm our expectations and their implications for the quantum dynamics. In addition, our observations on the branching ratio and the corresponding domain probabilities emphasize an important point -- strong coupling to the cavity does alter the dynamics even though a highly averaged or composite observable like reaction rate or branching ratio does not appear to change under VSC. 

We note that the model potential used here is low dimensional and one may justifiably ask whether the effects seen here would persist for more realistic PES corresponding to PTSB systems.  Answering this is challenging since obtaining sufficiently accurate PES for actual PTSB systems is not computationally feasible and on the fly  molecular dynamics simulations, as used in most of the dynamical studies earlier, are limited to the classical regime and constrained in terms of number and duration of the trajectories\cite{kuanjpca2024}. Furthermore, dissecting the  influence of the dozens of vibrational modes coupled to the reaction coordinate, potentially involved in a complex IVR dynamics, to identify the true VSC effects presents a rather difficult task at the moment. Interestingly, even researchers in the PTSB field have emphasized\cite{Harechemsci,chuangjctc,kramer2015reaction} the utility of constructing reduced dimensional models that would capture the essential dynamical features which influence the branching ratios. Thus, we believe that minimal models of the kind analyzed here are useful towards gaining initial insights into the mechanism of VSC modulated chemistry.

\begin{figure}[b]
    \centering
    \includegraphics[width=\columnwidth,height=8cm]{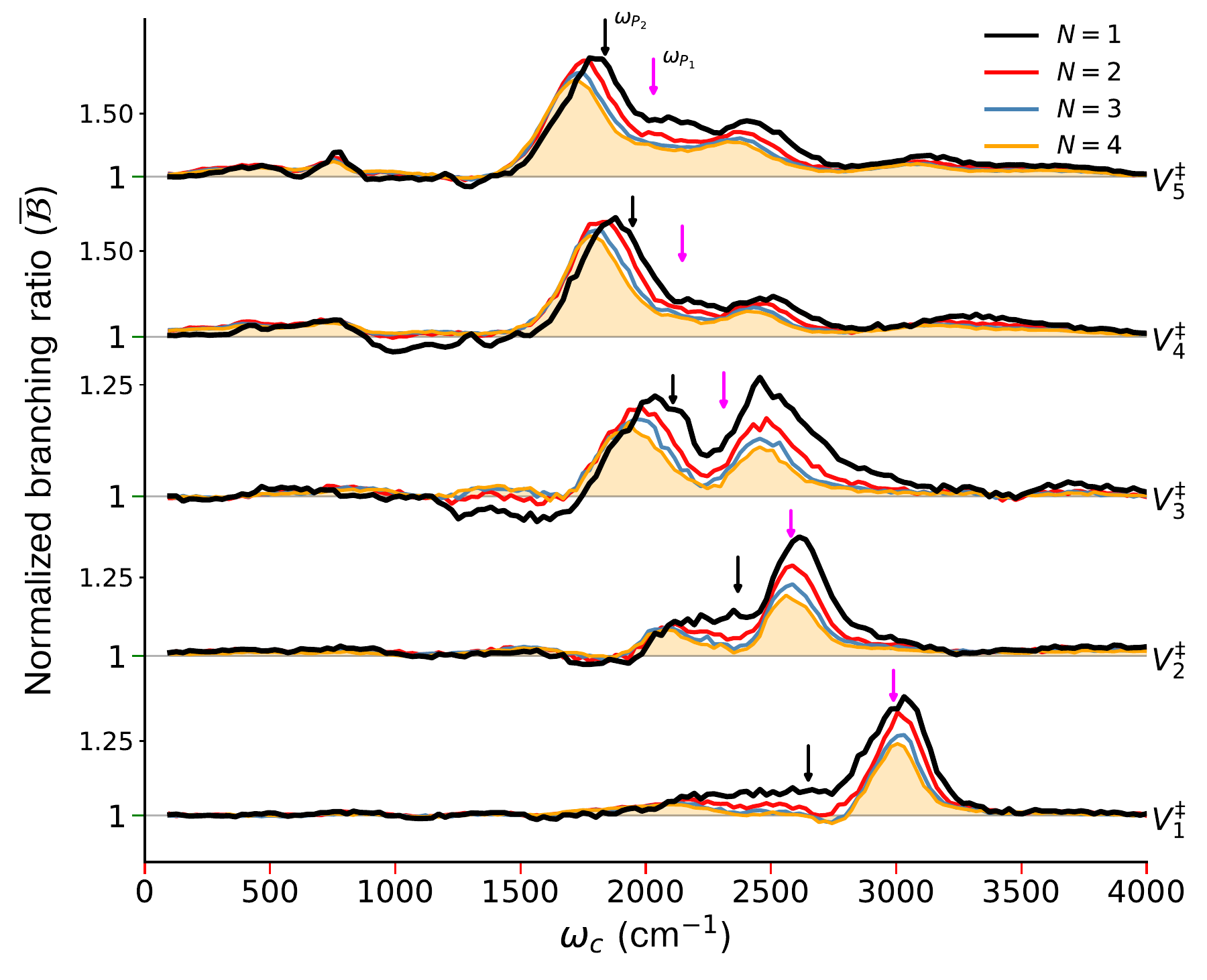}
    \caption{Classical normalized time averaged branching ratios shown for specific $V^{\ddagger}_{n}$ values for $N=1,\ldots,4$ number of molecules in the cavity. The coupling strength is scaled as $ \lambda_c = \frac{\lambda_0}{\sqrt{N}}$ with $\lambda_0 = 0.1$ au being the single molecule value. to keep the cavity-molecule interaction constant. Note that the initial energy for each molecule is fixed, as in the single molecule case, to be $\langle E_{S} \rangle = 0.01$ au. As in Fig.~\ref{fig:xcoupledBR}(c), the single molecule $N = 1$ harmonic product well frequencies are indicated by arrows.}
    \label{fig:4mol_Br}
    \end{figure}

 Although the present study focuses on a single-molecule reaction coupled to a single cavity mode, it provides a necessary starting point for exploring additional effects that arise under typical experimental conditions. These include the collective limit\cite{vega2025theoretical,perez2023simulating} involving $\sim 10^{6}$ molecules in the optical cavity, influence of cavity losses and solvent induced dissipation\cite{lindoy2023quantum,ahn2023modification,lindoy2024investigating,ke2024quantum,ying2024resonance}, and the limitations of considering only a single cavity mode.  At this stage the persistence of the single-molecule and single-mode  mechanism in the collective limit and in multimode lossy cavity environments is difficult to assess definitively. Nevertheless, as a preliminary result, in Fig.~\ref{fig:4mol_Br} we show the influence on the  branching ratio upon taking into account more than one molecule $N > 1$ in the cavity. For this calculation the $N$ molecular dipoles are taken to be aligned with the cavity polarization. Despite the fact that the results shown in Fig.~\ref{fig:4mol_Br} are classical, it is clear that the enhanced branching ratio peaks occur around the same cavity frequencies as in the single molecule case. In particular, based on recent studies\cite{lindoy2024investigating,rokaj2025,phusjctc2024}, we anticipate that for non-equilibrium and uncorrelated initial states the small $N$  effect on the branching ratio due to VSC should survive the $N \rightarrow \infty$ limit. More importantly, the phenomenon of switching of the cavity frequency with the changing well depth parameter persists in the few molecule case. Note that the collective coupling should not be
interpreted as requiring all molecules to pass through the transition-state
region synchronously. In our model, each molecule is coupled to the same cavity coordinate, and the light--matter interaction is present throughout the molecular system dynamics. The collective character therefore arises from the coupling of multiple molecular dipoles to the common cavity mode, rather than from simultaneous occupation of the transition-state region by all molecules. 
In particular, in the calculations the molecules  are not constrained
to cross the transition state synchronously. Each molecular trajectory evolves independently on its own molecular potential-energy surface, along its own reaction coordinate, while all molecules interact with the same cavity degree of freedom. The cavity therefore provides a common dynamical channel through which the molecular vibrational and photonic degrees of freedom can exchange energy. A molecule entering the post-transition-state region can consequently exchange energy with the common cavity mode even when the other molecules are at different locations along their respective reaction coordinates.
From this perspective, the role of the cavity in the collective case is better viewed as a shared dynamical mediator rather than as a mode that couples only to a transient transition-state population. The cavity-mediated energy exchange identified in the single-molecule case can therefore modify the post-transition state dynamics and, consequently, product selectivity in the collective case as well.
Moreover, inclusion of cavity loss and  the presence of multiple cavity modes\cite{ke2025harnessingmultimodeopticalstructure} may further enhance the VSC effect by offering additional channels for vibrational energy redistribution. 
 
 Therefore, the central physical insight emerging from this study — namely, that a properly tuned cavity mode can  act as an effective vibrational ``bath” that resonantly cools and stabilizes the system in a selected product well, is expected to hold under experimentally relevant conditions.   Nevertheless, a detailed study of the system within an open quantum approach is required to explore the parameter regimes wherein the observed effects are robust. Finally, we note that a recent paper\cite{Mondal2026} by Mondal, Vega, and Huo proposes that  the VSC induced effects can be explained by considering the many-body nature of the VSC. However, note that even in this new proposal only rate suppression is predicted, with the authors noting that the ``rate constant enhancements would likely depend on the details of the reaction mechanisms". Since dynamical effects in nonstatistical reactions like PTSB significantly influence\cite{schaeferorgmet2025,NielsenJACS2026} the reaction mechanism and selectivity, it remains to be seen if the many body viewpoint can be extended to multidimensional PESs exhibiting complex features, such as the one studied in this work.


\section*{Supplementary material}
The supplementary material includes the following: S$1$ gives details of the model potential surface used, S$2$ summarizes the numerical convergence studies performed, S$3$ shows additional examples of the branching ratio modulation under VSC for different parameter values, S$4$ provides more details regarding the various domain probabilities and the setting up of the grid for the VSC landscape, S$5$ shows the branching ratio in the zero photon case, S$6$ gives the equations used to compute average classical and quantum energies, and S$7$ provides a brief summary of the methodology employed for performing classical and quantum dynamics.

\section*{Data availability}

The data underlying this study are available at \href{https://github.com/kumaratul9766/PTSB-VSC.git}{https://github.com/kumaratul9766/PTSB-VSC.git} and the Supporting Information file provides details of the numerical techniques used and the various convergence studies.

\section*{Acknowledgements}
S.M. is grateful to the Ministry of Education, Government of India, for the Prime Minister Research Fellowship (PMRF) and the Fellowship for Academic and Research Excellence (FARE). A.K. thanks IIT Kanpur for the fellowship and the High Performance Computing (HPC) Center at IIT Kanpur for providing computing resources.

\bibliography{PTSBVSC} 

\end{document}


\maketitle

\begin{center}
\small
\textsuperscript{[a]}
Department of Chemistry, \\ Indian Institute of Technology Kanpur,
Uttar Pradesh 208016, India\\
\textsuperscript{*}E-mail: \href{mailto:srihari@iitk.ac.in}{srihari@iitk.ac.in}

\vspace{0.2cm}
\textsuperscript{+}These authors contributed equally.
\end{center}

\vspace{0.5cm}

\newcommand{\SIcontentsline}[2]{%
\noindent\textbf{\ref*{#1}.~\hyperref[#1]{#2}}
\leaders\hbox{.\kern0.5pt}\hfill%
\hyperref[#1]{\pageref*{#1}}%
\par\vspace{0.25cm}
}

\section*{\centering Table of Contents}

\vspace{0.5cm}

\SIcontentsline{sec:hamiltonian}{Details of the model potential energy surface}
\SIcontentsline{sec:convergence}{Numerical convergence studies}
\SIcontentsline{sec:branchingratio}{Branching ratio for different choice of parameters values}
\SIcontentsline{sec:Tavg_domain}{VSC landscape : time averaged domain probabilities}
\SIcontentsline{sec:zerophoton}{Zero photon case}
\SIcontentsline{sec:avgE}{Computing average energy components from the Pauli-Fierz Hamiltonian}
\SIcontentsline{sec:Method}{Methodology}
\SIcontentsline{sec:refs}{References}

\newpage

\section{Details of the model potential energy surface}
\label{sec:hamiltonian}

The two-dimensional PTSB model employed in this work is adapted from Tao et al.\cite{tao2023symmetric} and designed to provide a minimal yet flexible representation of post-transition state bifurcation dynamics while retaining clear physical interpretation. The potential energy surface is constructed as a sum of one-dimensional anharmonic contributions along the reaction coordinate $x$, a transverse coordinate 
$y$, and coupling terms that generate the bifurcation and valley–ridge inflection (VRI) features. The parameters are chosen to ensure the existence of a reactant minimum, an index-$1$ saddle (TS$_1$), a VRI region leading to two competing product wells, and a second index-$1$ saddle (TS$_2$) separating the two products.

\begin{figure}[H]
    \centering  
    \includegraphics[width=0.5\textwidth]{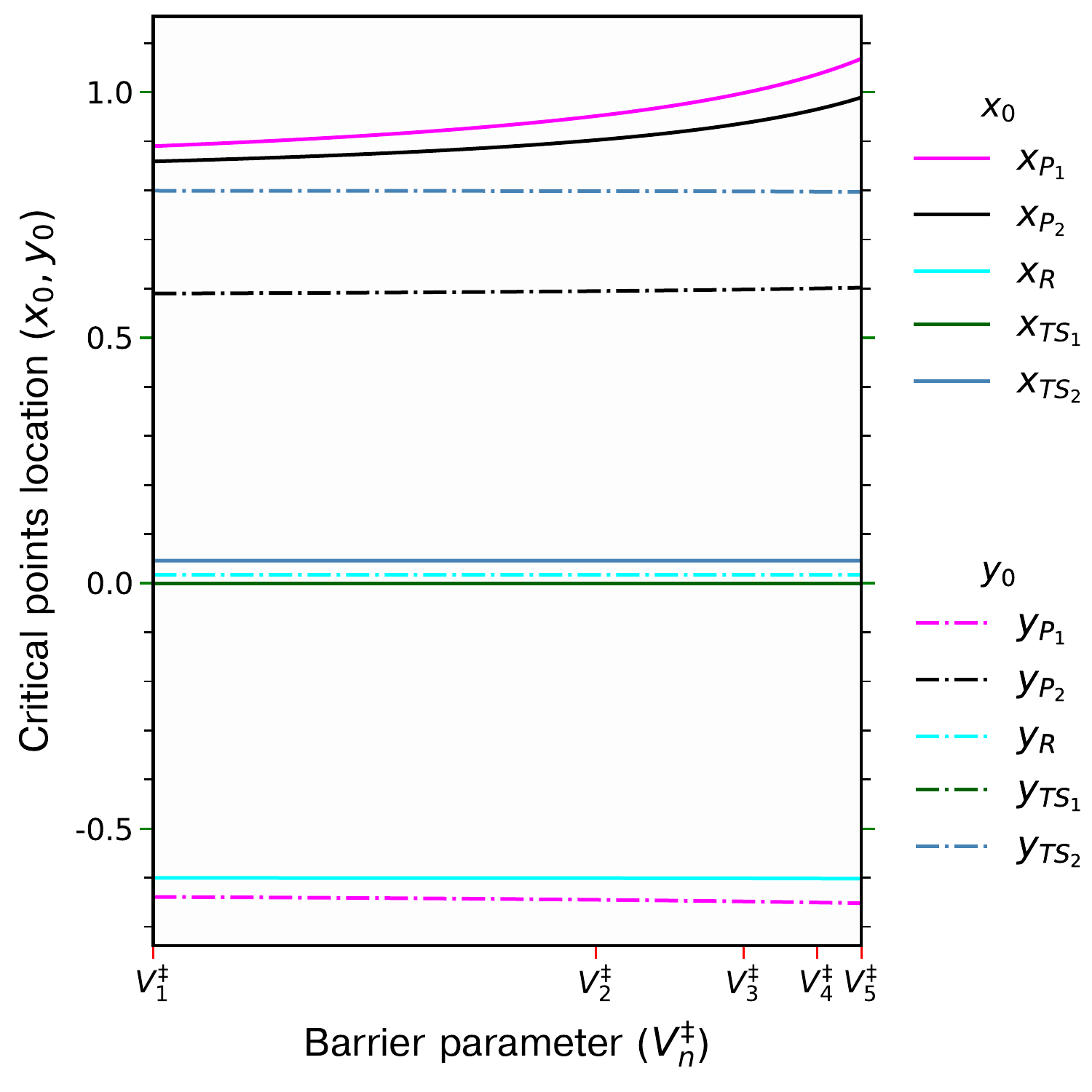}
    \caption{Variations in the critical points location with varying parameter $\overline{V}^{\ddagger} \equiv V^{\ddagger}_n/x_{s}^{4}$ with $V^{\ddagger}_n \equiv V^{\ddagger}/n$. }
    \label{fig:critical_points}
    \alttext{Line graph of the x and y coordinates of the five critical points against the barrier parameter. The two product x-coordinates rise gradually as the barrier parameter increases, while all of the remaining coordinates stay essentially flat across the whole range.}
    \end{figure}

The $x$-coordinate parameters  $A=1/2$, $B=2/3$, $C=12$ in the $x$-dependent part of the potential
\begin{equation*}
    U_x({x}) = \overline{V}^{\ddagger}\left[A x^{4} - B x_{s} x^{3} -C x_{s}^{2} x^{2}\right]
\end{equation*}
are fixed throughout the study to control the qualitative shape of the reaction pathway. Here, $\overline{V}^{\ddagger} \equiv V^{\ddagger}_n/x_{s}^{4}$ with $V^{\ddagger}_n \equiv V^{\ddagger}/n$ sets the depth of the wells, where $x_{s}=0.2$ sets the distance between R and TS$_{2}$. This choice ensures a sufficiently anharmonic landscape with a well-defined barrier at TS$_1$ and a smooth descent into the post-transition-state region. Fixing these parameters allows systematic exploration of how the overall energy scale controlled by $\overline{V}^{\ddagger}$-influences bifurcation dynamics without altering the underlying topology of the PES. Varying $V^{\ddagger}_n$ rescales the depths of the reactant and product wells as well as the barrier height, while preserving the relative positions of stationary points, mainly R, TS$_1$, and TS$_2$, (Fig. \ref{fig:critical_points}), while position of P$_1$ and P$_2$ change slightly. This makes $V^{\ddagger}_n$ a natural control parameter for probing how changes in well depth and curvature affect trajectory transport, resonance conditions, and cavity-induced selectivity. If $A$, $B$, and $C$ were varied instead, the qualitative shape of the PES-including location of stationary points, could change, complicating interpretation by mixing topological and energetic effects.

\textbf{Choice and scaling of $V^{\ddagger}_n$:} 
The barrier-scaling parameter $V^{\ddagger}$ used in the present work is adapted from the PTSB model introduced by Tao et al. \cite{tao2023symmetric}.
In that work, $V^{\ddagger}$ sets the overall energy scale of the potential, controlling both the depth of the reactant well and the energy separation between the first transition state (TS$_1$) and the post-transition-state region, while preserving the topology of the PTSB surface. 

For the purposes of the current study, the reference value of $V^{\ddagger}$ from \cite{tao2023symmetric} was scaled upward by a factor of 50 to $V^{\ddagger}= 3/1280$. This scaling was introduced to increase the absolute depth of the reactant and product wells and to ensure robust post-transition state trapping and well-defined product basins over the range of parameters investigated. Importantly, this uniform scaling does not alter the location or index of stationary points, nor does it modify the qualitative bifurcation topology of the potential energy surface; it only rescales energies and curvatures.\\
Starting from this scaled reference value $V^{\ddagger}$, it was then systematically reduced across a sequence of values $V^{\ddagger}_n$ in order to probe how progressive softening of the potential energy surface influences post-transition-state dynamics, cavity-induced resonances, and product selectivity. Because $V^{\ddagger}$ acts as a global energy prefactor in the model Hamiltonian, varying it provides a controlled way to tune well depths and barrier heights without changing the underlying PTSB structure. As a result, trends observed upon reducing $\overline{V}^{\ddagger}$ as $V^{\ddagger}_n$ can be directly attributed to energetic rescaling rather than changes in PES topology. The transverse potential
\begin{equation*}
    U_y({y}) = D y^{2} + F y^{4}
\end{equation*}
is chosen with positive $D=F=2/5$ to ensure confinement in the $y$-direction and to stabilize the product wells after bifurcation. These parameters set the stiffness of the transverse mode and therefore control the sensitivity of the system to lateral deviations near the VRI. Increasing $D$ or $F$ would suppress transverse excursions and reduce bifurcation sensitivity, whereas smaller values would enhance transverse instability and increase the likelihood of chaotic post-TS motion. In the present work, these parameters are fixed to isolate the effects of cavity coupling and barrier scaling.

\begin{figure}[H]
    \centering
    \includegraphics[width=0.5\textwidth]{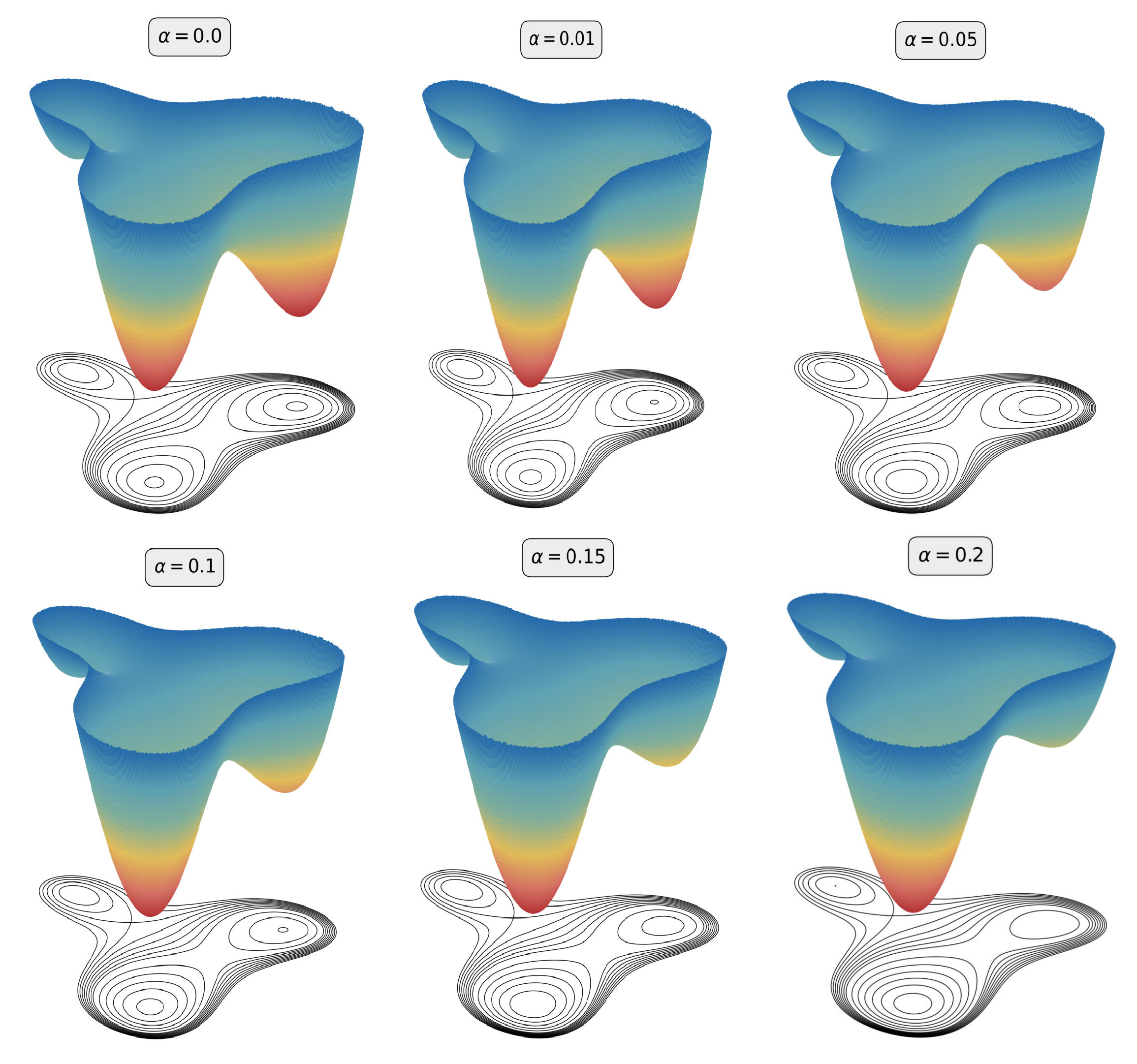}
    \caption{PTSB potential energy surfaces shown for different values of the asymmetry parameter $\alpha$ for the case of $V^{\ddagger}_2$.}
    \label{fig:varying_alpha}
    \alttext{Six three-dimensional surface plots, each with a contour map projected below it, one per asymmetry value. As the asymmetry parameter increases from zero to 0.2, one product well deepens while the other becomes progressively shallower and its contour rings shrink.}
    \end{figure}

The coupling potential
\begin{equation*}
    U_\alpha({x,y}) = G x^{2}y^{2} + Hx y^{2}(y^{2} -2) + \alpha xy
\end{equation*}
with  the coupling constants $G=1/2$, $H = 1$, and asymmetry parameter $\alpha$ introduces coupling between the reaction  and transverse coordinates.  The asymmetry parameter $\alpha$ breaks the symmetry between the two product channels and establishes a baseline preference for one product well over the other in the absence of coupling to the cavity. Keeping $\alpha$ fixed allows cavity-induced modifications of selectivity to be interpreted relative to a well-defined intrinsic bias. Larger values of $\alpha$ would dominate the branching behavior and reduce the relative influence of the cavity, while very small would produce nearly symmetric branching, obscuring cavity effects. For the present model, the overall PTSB topology, including the existence of a well-defined valley–ridge inflection (VRI) region separating the product wells, remains intact up to $\alpha \lesssim 0.2$.  The robustness of the PTSB topology for $\alpha \lesssim 0.2$ is illustrated in Fig.\ref{fig:varying_alpha}.

The nuclear mass is set to the carbon mass ($m=22034$ au) for both coordinates, providing a physically reasonable timescale for vibrational motion, with the model system of interest\cite{hare2017post}, while maintaining numerical stability in trajectory propagation. Changing the mass would rescale dynamical timescales and resonance conditions but would not qualitatively alter the PES topology.
    \begin{figure}[H]
    \centering  
    \includegraphics[width=0.5\textwidth]{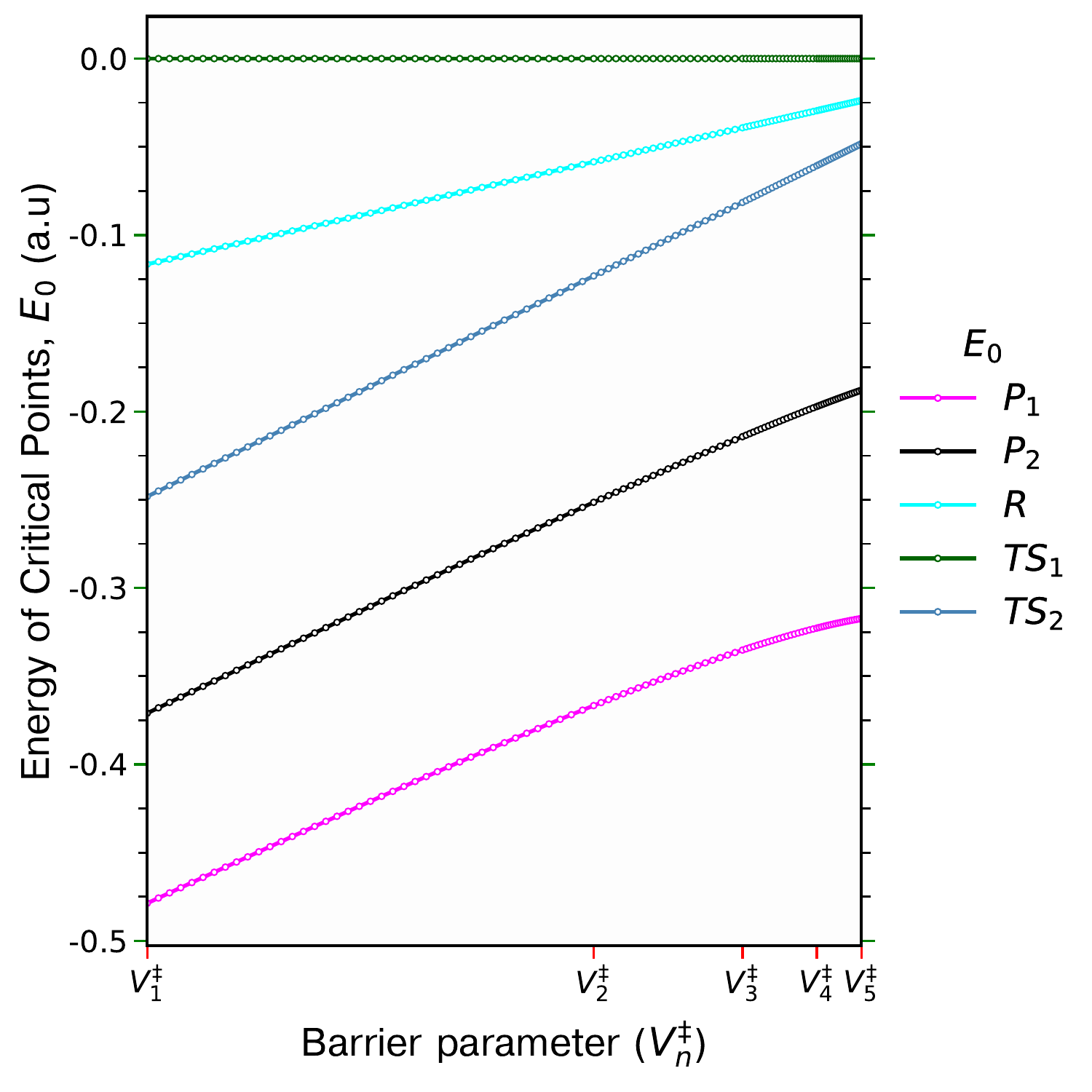}
    \caption{Variation of the products ($P_{1,2}$) and reactant ($R$) relevant well depths (in a.u) with $\overline{V}^{\ddagger}$. The energies of the two transition states $TS_{1,2}$ are also shown.}
    \label{fig:Energy_ofcriticalpoints}
    \alttext{Line graph of critical-point energies in atomic units against the barrier parameter. All four well and saddle energies rise roughly linearly toward the fixed first transition state energy, with the deeper product well lying lowest throughout.}
    \end{figure}
The cavity mode is modeled as a harmonic oscillator linearly coupled to the molecular dipole along the respective coordinate. The cavity-system coupling strength $\lambda_c$ is chosen to be in the VSC regime, ensuring that cavity effects are strong enough to influence dynamics. 

The specific numerical values chosen for the model parameters are not unique, and alternative parameter sets could be employed without qualitatively altering the conclusions of this study. The primary objective in constructing the present potential energy surface is to realize a well-defined post-transition state bifurcation (PTSB) topology, characterized by a reactant well, an index-1 transition state (TS$_1$), a valley–ridge inflection region, and two competing product wells separated by a second index-1 saddle (TS$_2$). As long as this critical-point structure is preserved, the essential dynamical features governing bifurcation and product selectivity remain unchanged. Variations in parameter values would mainly rescale energies, curvatures, or timescales\cite{tao2023symmetric,collins2013nonstatistical,KatsPRE2020}, but would not significantly modify the qualitative behavior of the system or the mechanisms explored in this work. The chosen parameters, therefore, serve to ensure a clear and controlled realization of a PTSB system with well-resolved stationary points, allowing systematic investigation of cavity-induced effects on post-transition state dynamics.

\subsection{Definition of the VRI}
A valley–ridge inflection (VRI) point is a non-stationary point on a potential energy surface (PES) at which the local character of the surface changes from a valley to a ridge (or vice versa) along a direction orthogonal to the reaction coordinate. Unlike minima or saddle points, the gradient of the potential does not vanish at a VRI point; instead, the defining feature is a change in curvature topology.\\
Mathematically, a VRI point is characterized by the following two conditions,
\begin{enumerate}
    \item \textbf{Vanishing curvature in one direction:} \\
    \\
    The Hessian matrix $\mathbf{H}$ of the potential has at least one zero eigenvalue,
    \begin{equation*}
    \det(\mathbf{H}) = 0
    \end{equation*}
where $\mathbf{H}$ is the Hessian matrix. This indicates an inflection in curvature.
\item \textbf{Orthogonality between gradient and zero-curvature direction:} \\

The gradient vector $\mathbf{g}=\nabla V$ is orthogonal to the eigenvector $\mathbf{v}_0$ corresponding to the zero eigenvalue of $\mathbf{H}$,
    \begin{equation*}
    \mathbf{g} \cdot \mathbf{v}_0 = 0
    \end{equation*}
\end{enumerate}
Together, these conditions distinguish a VRI point from a conventional saddle point: the PES is flat along one direction but still experiences a net force.
In practice, VRI point can be located without explicit diagonalization of the Hessian by identifying the simultaneous solutions of two scalar conditions as the intersection of the zero contours of

\begin{equation*}
\mathbf{g}^{\mathsf T} \operatorname{adj}[\mathbf{H}] \mathbf{g} = 0 \quad, 
\qquad
\det(\mathbf{H}) = 0
\end{equation*}

where $\operatorname{adj}[\mathbf{H}] = \det(\mathbf{H}) \mathbf{H}^{-1}$
is the adjugate of the Hessian.

\textbf{Significance of the VRI in the PTSB systems}

In post-transition state bifurcation (PTSB) reactions, the VRI region plays a central role in controlling product selectivity. After crossing the first transition state, reactive trajectories encounter the VRI region, where the PES no longer guides motion uniquely along a single downhill path. Instead, small perturbations of the initial conditions, intramolecular vibrational energy redistribution, or external coupling such as an optical cavity can divert trajectories toward different product channels. Consequently, the VRI region acts as a dynamical gateway that mediates branching behavior beyond the transition state.

For the present PTSB model, the VRI point is located at 
$(x,y)$=$(0.21,0.005)$ for the parameter set corresponding to $V^{\ddagger}_1$ with an associated potential energy of $E=-0.0313$ au. Upon reducing 
$V^{\ddagger}_n$ (increasing $n$), the position of the VRI point in configuration space remains unchanged, while its energy decreases systematically in accordance with the overall rescaling of the potential.

\subsection{Average System Energy associated with the initial wavepacket}

    \begin{figure}[H]
    \centering  
    \includegraphics[width=0.7\textwidth]{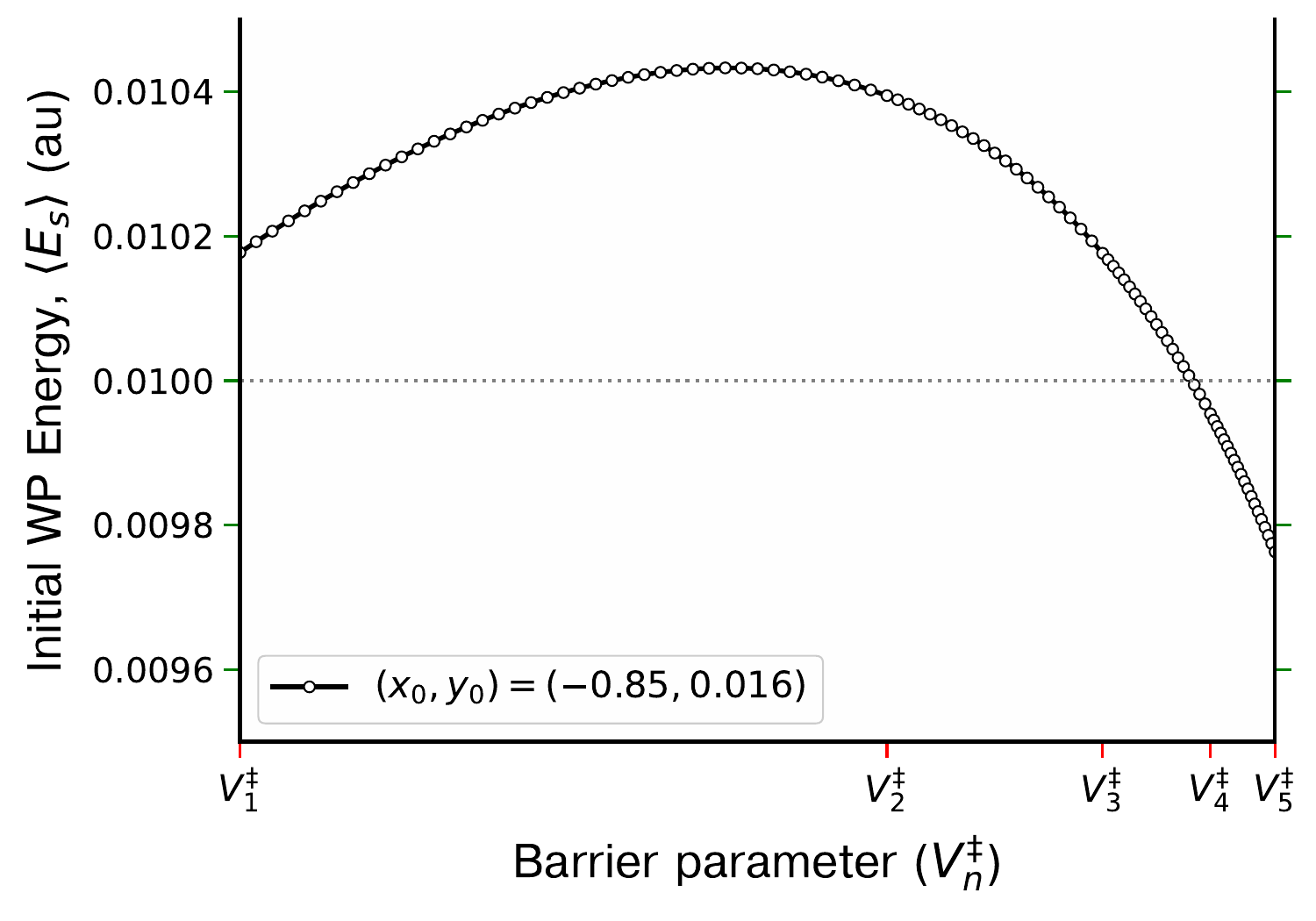}
    \caption{Variation of the initial average system energy, $\langle E_{S}\rangle = \langle \psi|H_{M}|\psi\rangle$ (in au) with changing  $\overline{V}^{\ddagger}$, for an initial wavepacket centered around the reactant well at $(x_{0},y_{0}) = (-0.85,0.016)$ au.}
    \label{fig:InitialWPenergy}
    \alttext{Line graph of the initial wavepacket average energy against the barrier parameter. }
    \end{figure}

\subsection{Time Averaged Bare Branching ratio (without cavity)}

    \begin{figure}[H]
    \centering  
    \includegraphics[width=0.7\textwidth]{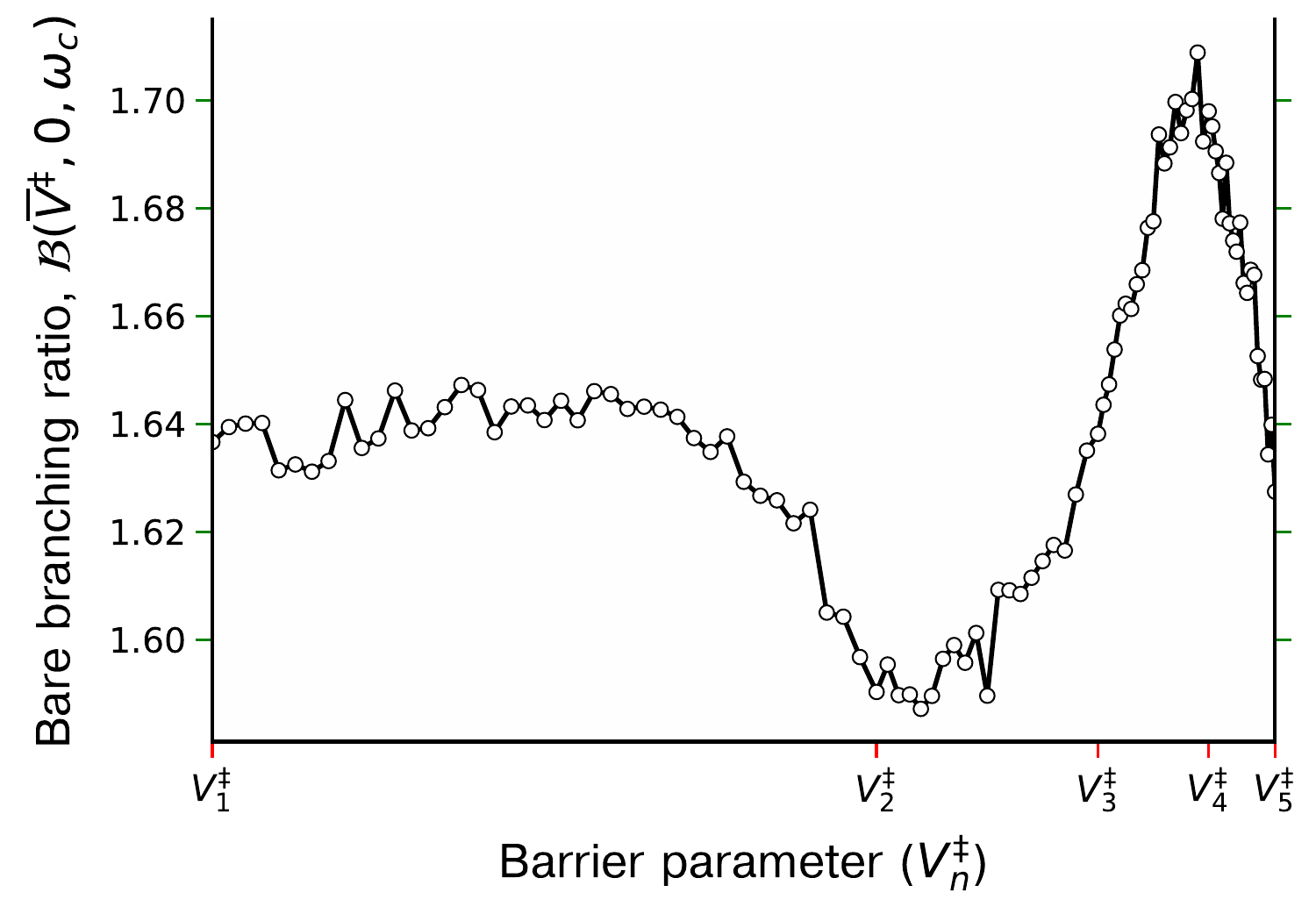}
    \caption{Variation of time averaged bare branching ratio (without cavity coupling, $\lambda_{c} = 0$) with variation in $\overline{V}^{\ddagger}$ for initial wavepackets  centered around the reactant well at $(x_{0},y_{0}) = (-0.85,0.016)$ au with
    initial system energies $\langle E_{S} \rangle \approx 0.01$ au.}
    \label{fig:BR_withoutcavity}
    \alttext{Line graph of the cavity-free branching ratio against the barrier parameter.}
    \end{figure}
    
\section{Numerical convergence}

\begin{figure}[H]
    \centering  
    \includegraphics[width=0.9\textwidth]{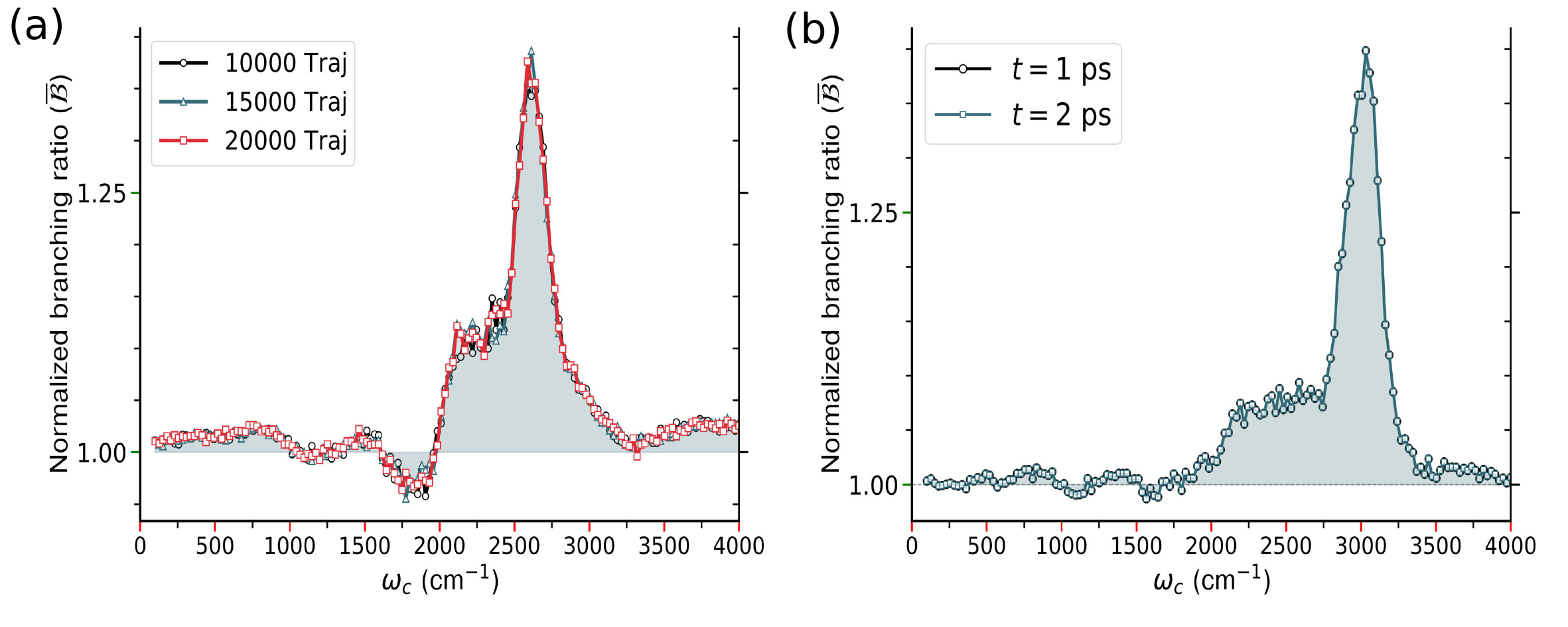}
    \caption{Branching ratio plots for the PTSB system under x-mode cavity coupling as a function of the cavity frequency $\omega_c$ (a) Convergence with respect to ensemble size for the case of  $V^{\ddagger}_2$ at $\langle E_s \rangle = 0.01$ au. The  grey baseline shows the normalized branching ratio for the uncoupled case. (b) Convergence with respect to total integration time for the case of  $V^{\ddagger}_1$ at $\langle E_s \rangle = 0.01$ au.}
    \label{fig:numconv_xcoupledBR}
    \alttext{Two line graphs of normalized branching ratio against cavity frequency. In the first, three ensemble sizes give overlapping curves; in the second, two total integration times also overlap, so peak position and height are converged in both tests.}
    \end{figure}

To ensure the robustness of the results, we performed systematic convergence tests with respect to both the total number of trajectories and the total integration time. Figure. \ref{fig:numconv_xcoupledBR}(a) shows the normalized branching ratio as a function of cavity frequency for three different ensemble sizes  for the representative case of $V^{\ddagger}_1$ at an average ensemble energy $\langle E_s \rangle = 0.01$ au. The resonance features and peak intensities are seen to be well converged at $20000$ trajectories, with no significant changes upon further increasing the ensemble size.

\label{sec:convergence}
   
Convergence with respect to the total propagation time, utilized to compute the average branching ratios, is shown in Fig. \ref{fig:numconv_xcoupledBR}(b).  Thus, based on these convergence tests, all classical simulations reported in the main text and supplementary information are performed using $20000$ trajectories and for a total integration time of $1$ ps.

    \begin{figure}[H]
    \centering  
    \includegraphics[width=0.85\textwidth]{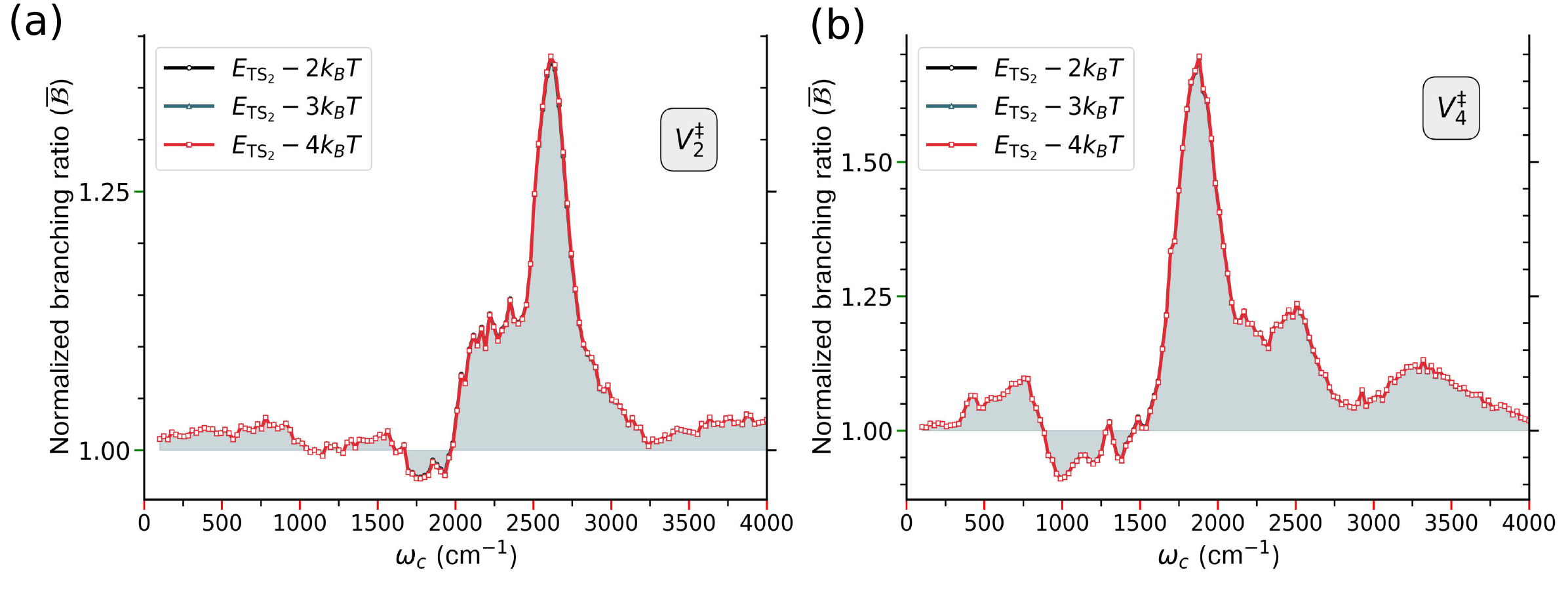}
    \caption{Branching ratios for the PTSB system under $x$-mode cavity coupling with the cavity frequency $\omega_c$ for different choices of the domain energy cutoff criteria. The  grey baseline shows the normalized branching ratio for the uncoupled case.}
    \label{fig:numconv_xcoupledBR_withdomain}
    \alttext{Two line graphs of normalized branching ratio against cavity frequency, one per barrier parameter. Three different domain energy cutoff criteria produce curves that overlap almost exactly, leaving both the resonance peak positions and their heights unchanged.}
    \end{figure}

\section{Branching ratio for different choice of parameter values}
\label{sec:branchingratio}

\subsection{Different values of initial ensemble-averaged energies \texorpdfstring{$\langle E_S\rangle$} {<Es>} and system-cavity coupling strengths \texorpdfstring{$\lambda_c$} {λc}}

\begin{figure}[H]
    \centering  
    \includegraphics[width=0.6\textwidth]{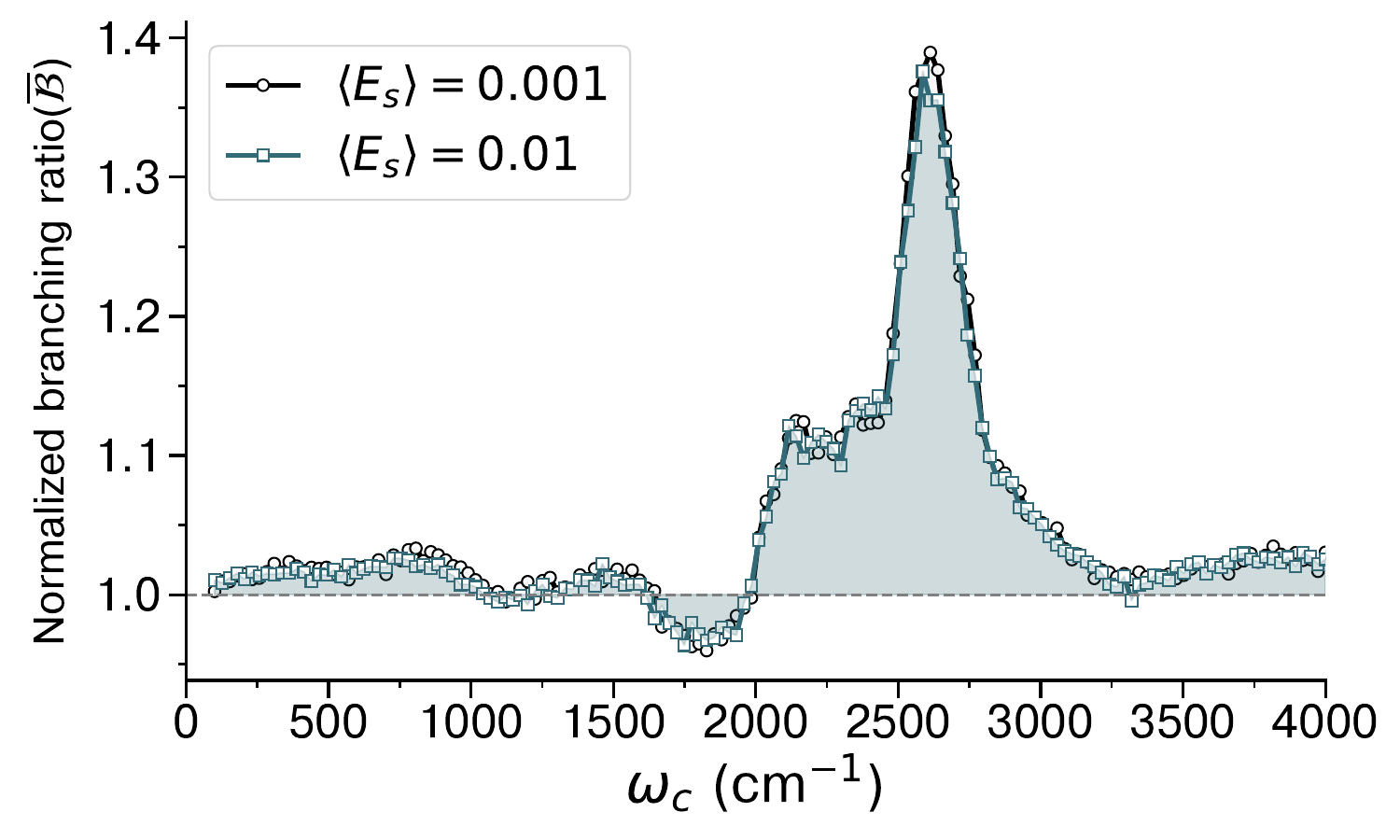}
    \caption{Branching ratios for the PTSB system under $x$-mode cavity coupling for two different initial average energy of ensemble $\langle E_s \rangle$ with fixed parameter $V^{\ddagger}_2$.}
    \label{fig:varyE_xcoupledBR}
    \alttext{Line graph of normalized branching ratio against cavity frequency for two initial ensemble energies. Both curves produce qualitatively the same result.}
    \end{figure}

In addition to numerical convergence, we examined the sensitivity of the branching ratio results to variations in the average ensemble energy $\langle E_S \rangle$ and the light–matter coupling strength $\lambda_c$. Fig. \ref{fig:varyE_xcoupledBR} compares the normalized branching ratio for two different initial ensemble energies. The overall features, including the position of the resonance peak and its enhancement relative to the uncoupled baseline, remain essentially unchanged. This indicates that the cavity-induced modulation of the branching ratio is not strongly dependent on the precise choice of initial ensemble energy, and reflects a robust dynamical effect rather than a fine-tuned energetic condition.

    \begin{figure}[H]
    \centering  
    \includegraphics[width=1.0\textwidth]{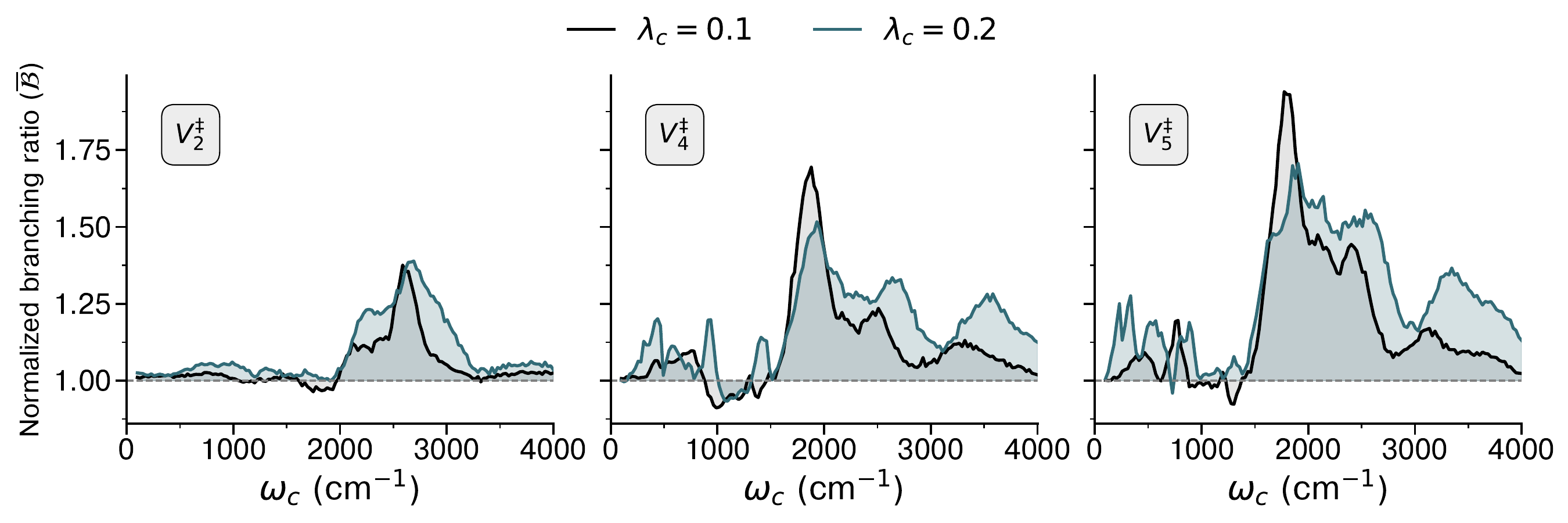}
    \caption{Branching ratios for the PTSB system under $x$-mode cavity coupling for different light-matter coupling strength $\lambda_c$  with fixed values of $\langle E_S \rangle = 0.01$ au and $\alpha = 0.1$.}
    \label{fig:varyeta_withVdag}
    \alttext{Three line graphs of normalized branching ratio against cavity frequency, one per barrier parameter. Doubling the coupling strength broadens each profile, but qualitatively preserves the branching ratio feature.}
    \end{figure} 
    
Figure. \ref{fig:varyeta_withVdag} shows the effect of increasing the light–matter coupling strength. While the resonance position remains similar, a noticeable broadening of the resonance feature is observed at larger $\lambda_c$. This broadening can be attributed to stronger hybridization between the molecular $x$-mode and the cavity mode, which effectively increases the range of frequencies over which energy exchange between the system and cavity is efficient. Importantly, the qualitative behavior of the branching ratio remains unchanged, confirming that the resonance mechanism persists even under stronger coupling, albeit with a broader response.

\subsection{Results for different values of the asymmetry parameter \texorpdfstring{$\alpha$}{α}}

\begin{figure}[H]
    \centering  
    \includegraphics[width=1.0\textwidth]{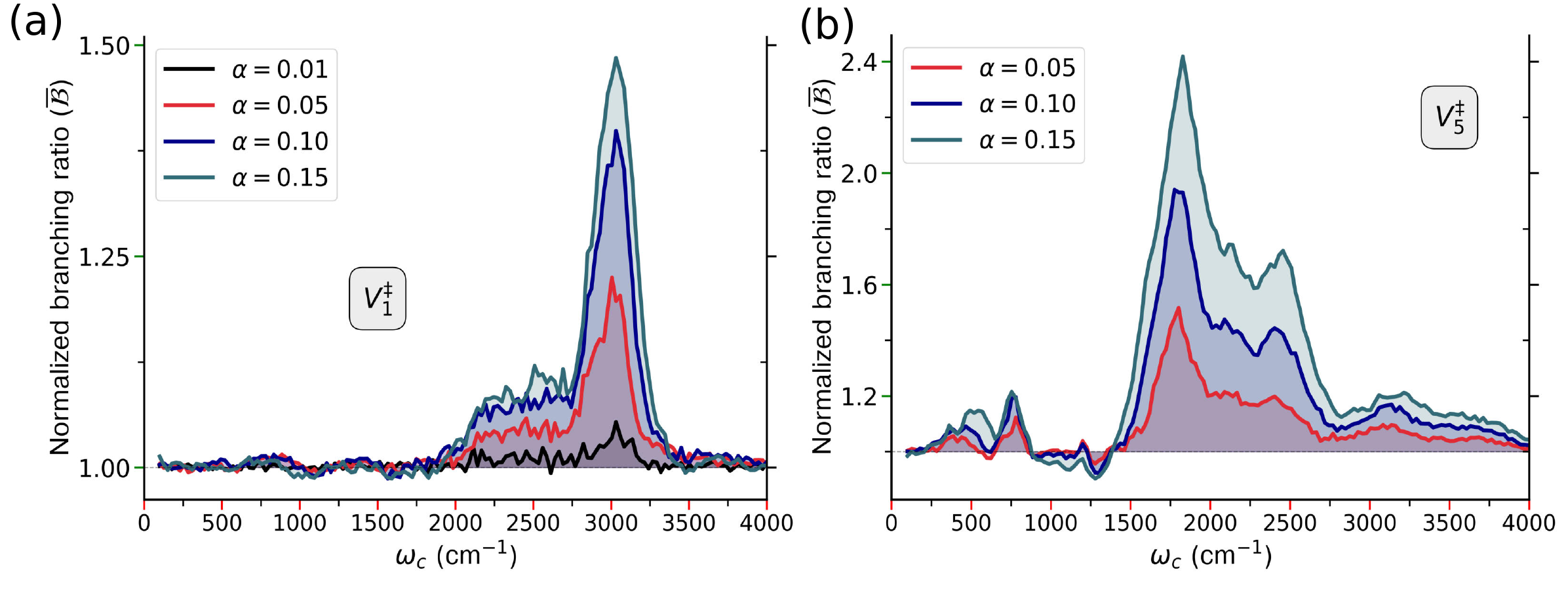}
    \caption{Branching ratios for the PTSB system under $x$-mode cavity coupling  for different $\alpha$ with fixed $\langle E_s \rangle = 0.01$ au and $\lambda_c = 0.1$.}
    \label{fig:varyalpha_xcoupled}
    \alttext{Two line graphs of normalized branching ratio against cavity frequency for several asymmetry values. Peak height grows steadily with the asymmetry parameter, reaching about 1.5 in the first panel and about 2.4 in the second, with no shift in peak frequency.}
    \end{figure}

We also examined the sensitivity of the cavity-modified branching ratio to variations in the asymmetry parameter $\alpha$, which controls the intrinsic bias between the two product wells. Figure.\ref{fig:varyalpha_xcoupled} shows the normalized branching ratio as a function of cavity frequency for several values of $\alpha$ in the $x$-coupled case. Across the range of asymmetries considered, the qualitative resonance features remain unchanged: the position of the dominant resonance peak and its overall frequency dependence are preserved. This indicates that the cavity-induced resonance mechanism is robust with respect to changes in the intrinsic product asymmetry.

As $\alpha$ is increased, however, a systematic increase in the peak intensity is observed. This behavior is expected, since a larger $\alpha$ strengthens the underlying preference for the dominant product well, thereby amplifying the effect of cavity-induced modulation when the system is driven near resonance. Importantly, the persistence of the same resonance structure across different $\alpha$ values further supports the interpretation that the observed branching ratio features arise from resonant coupling to the system vibrational modes, rather than due to the intrinsic asymmetry of the product wells. 

\subsection{Frequency-dependent scaling of the light--matter coupling}
\label{sec:lambda_scaling}

In the main manuscript, we use a fixed light--matter coupling strength  $\lambda_c=0.1$~a.u. while scanning the cavity frequency $\omega_c$. The dimensionless coupling parameter is defined as
\begin{equation}
    \eta \equiv \frac{\Omega_R}{2\omega_c},
\end{equation}
where $\Omega_R$ is the Rabi splitting. Since $\eta$ depends explicitly on the cavity frequency for fixed $\lambda_c$, it is not strictly constant over the entire cavity-frequency range considered in the calculations. The value $\eta\simeq0.025$  quoted in the main text that corresponds to $\Omega_R \sim 100$ cm$^{-1}$ should therefore be understood as an approximate characterization of the coupling regime around the relevant molecular frequencies.

To examine whether the frequency dependence of $\eta$ influences the cavity-induced branching-ratio features, we additionally performed calculations in which $\lambda_c$ was scaled with the cavity frequency so as
to approximately maintain a constant $\eta$.

The scaling follows directly from the form of the light--matter interaction in the Pauli--Fierz Hamiltonian. For the bilinear interaction
\begin{equation}
    H_{\mathrm{int}} = \omega_c\lambda_c q_c\mu(x),
\end{equation}
the harmonic cavity coordinate can be written in terms of the creation and annihilation operators as
\begin{equation}
    \hat q_c = \frac{1}{\sqrt{2\omega_c}} \left(\hat a+\hat a^\dagger\right),
\end{equation}
and consequently
\begin{equation}
    \langle 1|\hat q_c|0\rangle =
    \frac{1}{\sqrt{2\omega_c}}.
\end{equation}
Thus, for a fixed molecular transition dipole, the light--matter coupling matrix element scales as
\begin{equation}
    g \propto
    \omega_c\lambda_c
    \langle 1|q_c|0\rangle
    \propto
    \lambda_c\sqrt{\omega_c}.
\end{equation}
Since the Rabi splitting satisfies $\Omega_R\simeq 2g$, the dimensionless coupling parameter scales as
\begin{equation}
    \eta
    =
    \frac{\Omega_R}{2\omega_c}
    \propto
    \frac{\lambda_c}{\sqrt{\omega_c}}.
\end{equation}
Therefore, maintaining approximately constant $\eta$ requires
\begin{equation}
    \frac{\lambda_c(\omega_c)}{\sqrt{\omega_c}}
    =
    \frac{\lambda_{c,0}}{\sqrt{\omega_{c,0}}},
\end{equation}
or equivalently
\begin{equation}
    \boxed{
    \lambda_c(\omega_c)
    =
    \lambda_{c,0}
    \sqrt{\frac{\omega_c}{\omega_{c,0}}}
    } ,
    \label{eq:lambda_scaling}
\end{equation}
where $\lambda_{c,0}$ is the reference coupling at the reference cavity
frequency $\omega_{c,0}$.
\begin{figure}[H]
    \centering
    \includegraphics[width=0.65\textwidth]{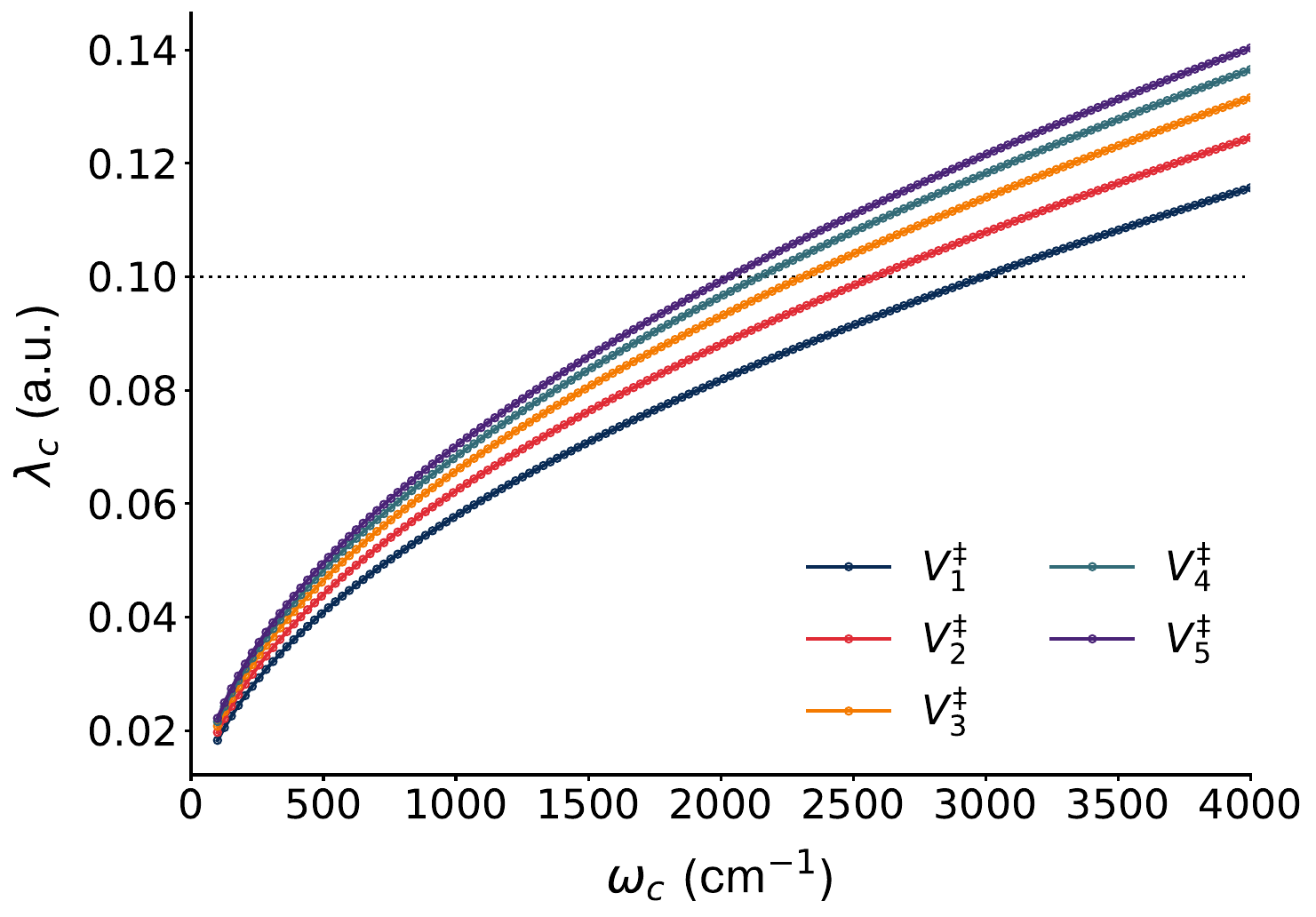}
    \caption{Frequency-dependent light--matter coupling strength used
    to maintain approximately constant dimensionless coupling $\eta$.
    The cavity coupling strength $\lambda_c$ is varied with cavity frequency
    according to
    $\lambda_c(\omega_c)=\lambda_{c,0}\sqrt{\omega_c/\omega_{c,0}}$,
    with $\lambda_{c,0}=0.1$ at the reference frequency $\omega_{c,0}$.
    Results are shown for the five barrier parameters
    $V_n^{\ddagger}$, $n=1,\ldots,5$.}
    \label{fig:lambda_scan}
    \alttext{Line graph of the frequency-dependent coupling strength against cavity frequency for five barrier parameter values. All five curves rise steeply at low frequency and then flatten progressively with cavity frequency, crossing the constant reference value of 0.1 atomic units between roughly 2000 and 3000 cm$^{-1}$.}
\end{figure}
Figure~\ref{fig:lambda_scan} shows the frequency-dependent values of $\lambda_c$ obtained from Eq.~\eqref{eq:lambda_scaling} for the five values of the barrier parameter, $V_n^{\ddagger}$, considered in this work. The required coupling strength increases systematically with increasing cavity
frequency, as expected from the square-root scaling in Eq.~\eqref{eq:lambda_scaling}.

\begin{figure}[H]
    \centering
    \includegraphics[width=0.8\textwidth]{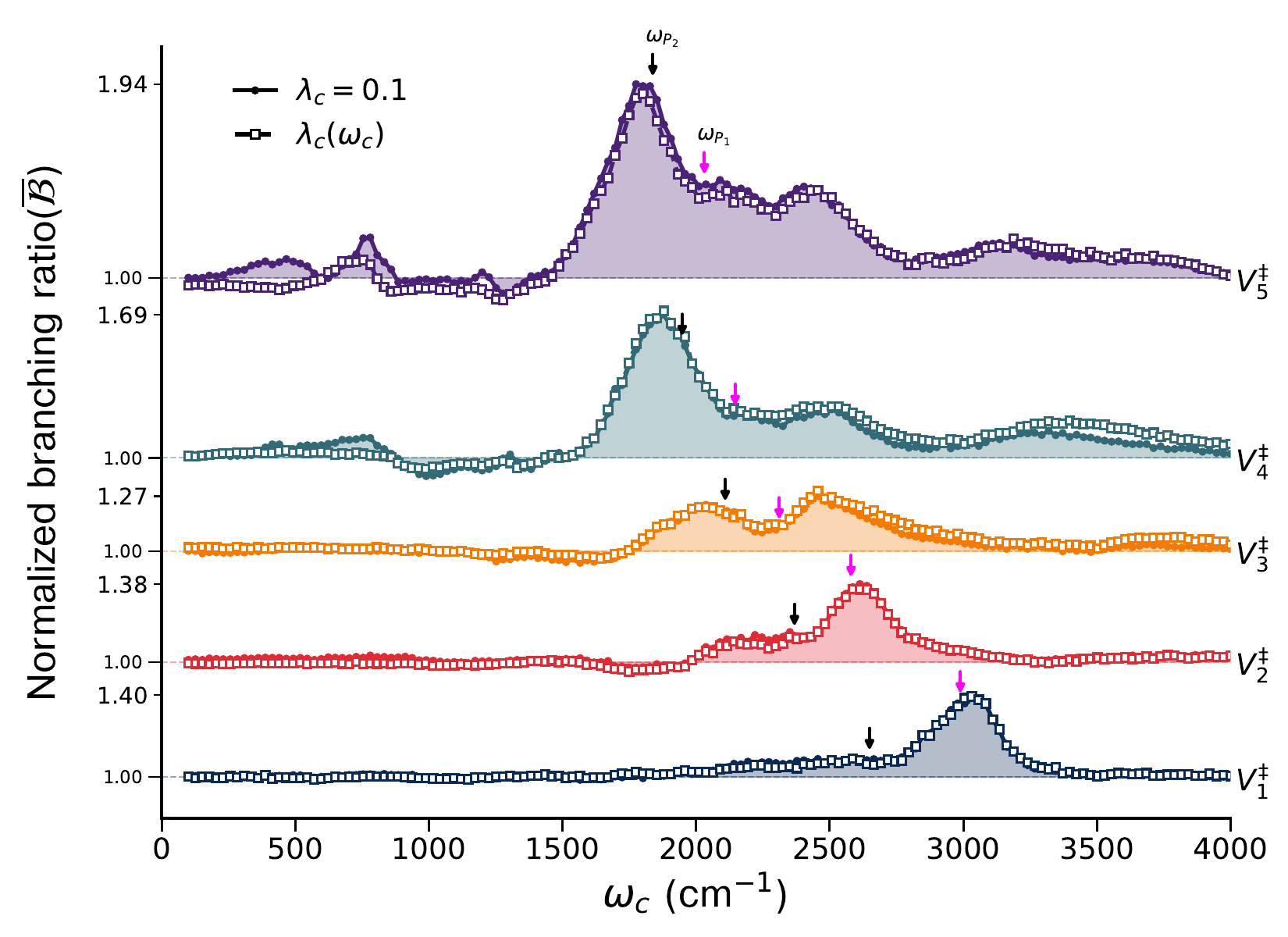}
    \caption{\textbf{Comparison of the normalized branching ratio obtained
    using fixed and frequency-dependent cavity coupling strengths.}
    Results are shown for $V_1^{\ddagger}$--$V_5^{\ddagger}$ as a function
    of the cavity frequency $\omega_c$. Solid curves with filled markers
    correspond to the original calculations with fixed $\lambda_c=0.1$,
    whereas curves with open-square markers correspond to the
    frequency-dependent scaling
    $\lambda_c(\omega_c)=\lambda_{c,0}
    \sqrt{\omega_c/\omega_{c,0}}$.}
    \label{fig:BR_lambda_scan}
    \alttext{Stacked line graphs of normalized branching ratio against cavity frequency for constant and frequency-dependent coupling, at five barrier parameter values. The two schemes trace nearly the same peaks}
    
We next compare the normalized branching ratios obtained with the original fixed coupling $\lambda_c=0.1$ and with the frequency-dependent scaling of Eq.~\eqref{eq:lambda_scaling}. The comparison is shown in
Fig.~\ref{fig:BR_lambda_scan} for all five values of $V_n^{\ddagger}$.

\end{figure}

The comparison demonstrates that the principal frequency-dependent features of the cavity-induced branching-ratio response are preserved when the frequency dependence of the coupling strength is explicitly taken into
account. In particular, the systematic evolution of the resonance structure across the sequence $V_1^{\ddagger}$--$V_5^{\ddagger}$, including the pronounced enhancement for $V_4^{\ddagger}$ and $V_5^{\ddagger}$, remains evident in the scaled-coupling calculations. The resonance features remain
associated with the same molecular dynamical frequency scales, although their intensities can change quantitatively, particularly for the softer potential-energy surfaces at smaller $V^{\ddagger}$.

These results indicate that the enhanced cavity response observed upon reducing $V^{\ddagger}$ cannot be attributed solely to the frequency dependence of the dimensionless coupling parameter resulting from keeping
$\lambda_c$ fixed. Rather, the characteristic frequency switching and the associated enhancement of product selectivity persist when the coupling is rescaled to approximately maintain a constant $\eta$. Thus, the principal qualitative conclusions of the present work are robust to this
frequency-dependent scaling of the light--matter coupling.

\section{VSC landscape: time averaged domain probabilities}
\label{sec:Tavg_domain}

The three-dimensional surface of the normalized branching ratio as a function of cavity frequency $\omega_c$ and $V^{\ddagger}_n$ was constructed on an effective $311 \times 100$ grid in the $(\omega_c,V^{\ddagger}_n)$ plane. 

    \begin{figure}[H]
    \centering  
    \includegraphics[width=0.7\textwidth]{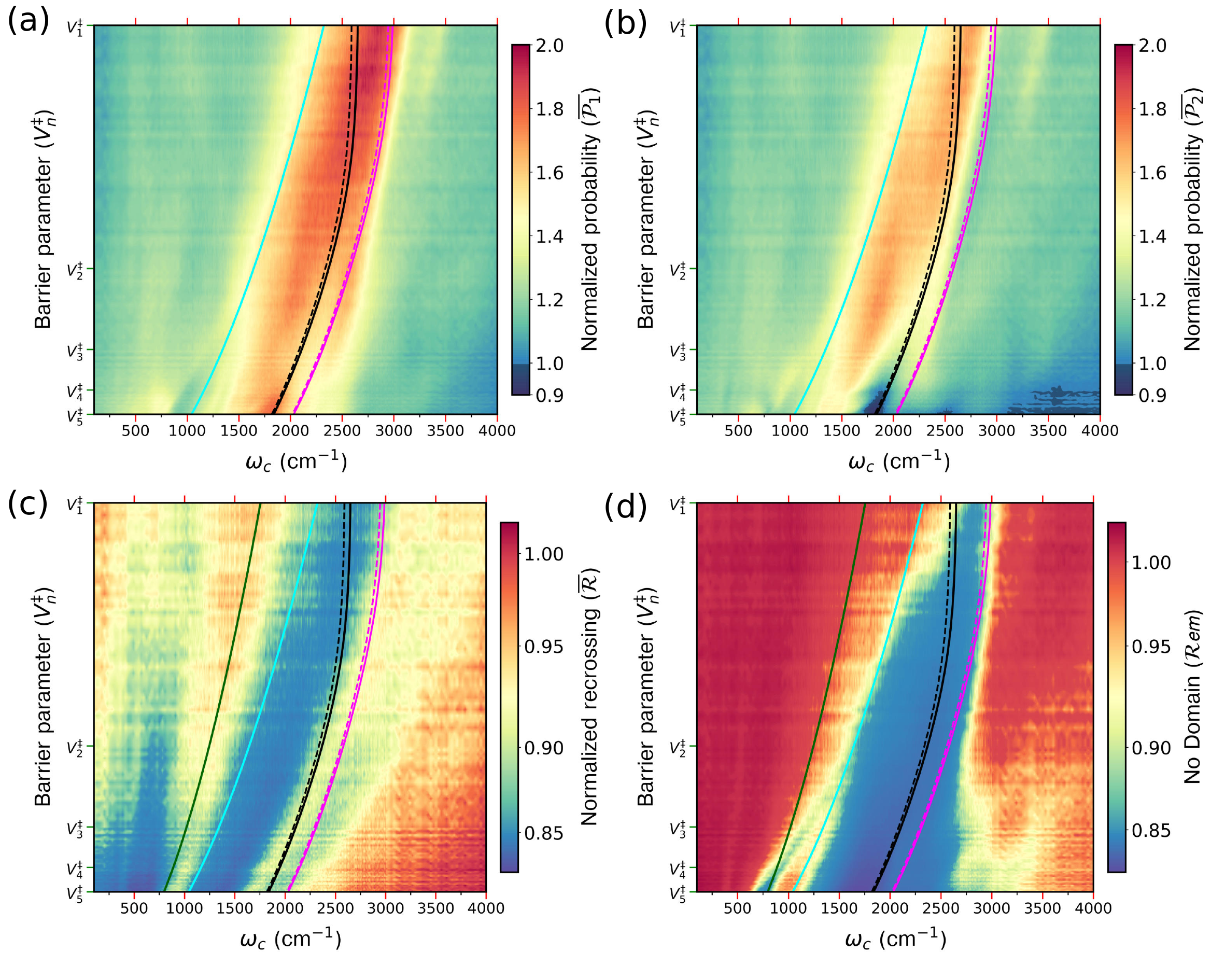}
    \caption{ Projection of normalized time-averaged domain probabilities corresponding to Fig. $3$(a) and (b) in main text, over the 2-D plane of $V^{\ddagger}_n$ and $\omega_c$ (cm$^{-1}$). Here solid and dotted lines denotes the associated harmonic frequencies along the reaction coordinate $x$ and fundamental frequencies in two product wells, respectively as shown in main text (refer Fig.$2$(a); color coded). Shown for the (a)  Normalized time-averaged domain probability $\overline{\cal{P}}_{P_{1}}$. (b) Normalized time-averaged domain probability $\overline{\cal{P}}_{P_{2}}$. (c) Recrossed normalized time-averaged domain probability $\overline{\cal{R}}$. (d) Case when the population has entered neither of the three domains.}
    \label{fig:domain_prob3d}
    \alttext{Four heat maps of normalized domain probability over cavity frequency and barrier parameter, overlaid with mode-frequency curves. The first two show product probability ridges that track the overlaid curves; the last two show matching troughs in recrossing and in undecided population.}
    \end{figure}
    
         The sampling along $V^{\ddagger}$ was intentionally chosen to be non-uniform in order to maintain adequate resolution across all barrier-scaling regimes. Specifically, 40 grid points were used between $V^{\ddagger}_1$ and $V^{\ddagger}_2$, followed by 20 points each in the intervals $V^{\ddagger}_2 \rightarrow V^{\ddagger}_3$, $V^{\ddagger}_3 \rightarrow V^{\ddagger}_4$, and $V^{\ddagger}_4 \rightarrow V^{\ddagger}_5$. This piecewise refinement was adopted because a uniformly spaced grid would have led to significant undersampling in the low $V^{\ddagger}_n$ regime. Along the cavity-frequency axis, an initial coarse scan was performed over $100 \leq \omega_c \leq 4000~\mathrm{cm}^{-1}$ using 150 grid points. Based on the resonance structure identified in this scan, the frequency range $1000 \leq \omega_c \leq 3000~\mathrm{cm}^{-1}$ was subsequently refined with an additional 161 points at a uniform spacing of 12.5~$\mathrm{cm}^{-1}$. The final three-dimensional surface, therefore, combines global sampling with targeted refinement in the physically relevant frequency regime.
         
    \begin{figure}[H]
    \centering  
    \includegraphics[width=1.0\textwidth,height=10.0cm]{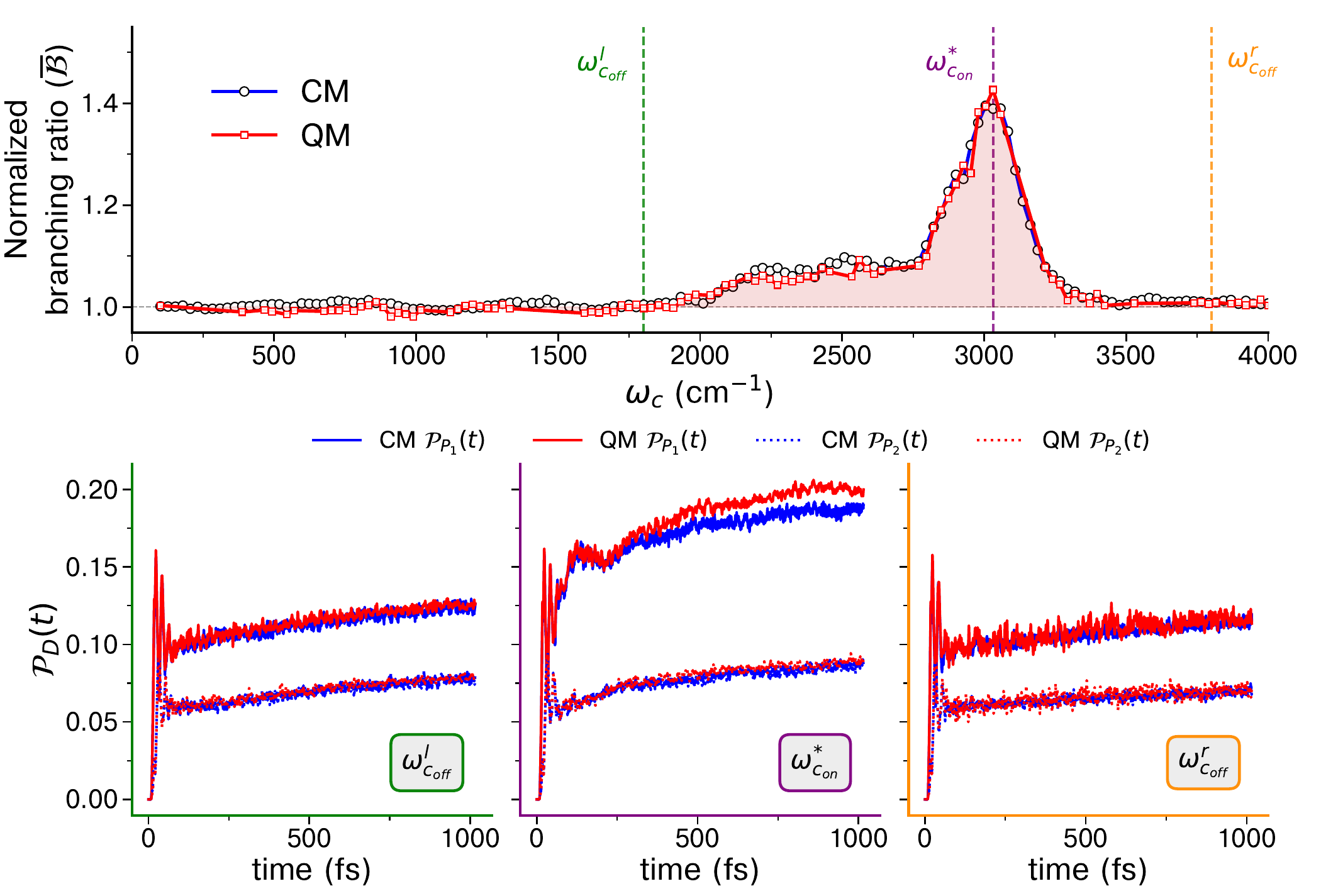}
    \caption{ Variation of normalized branching ratio with cavity frequency for the case of $V_1^{\ddagger}$ and the time-resolved product-domain probabilities $\mathcal{P}_D(t)$ at  on and off resonant conditions corresponding to that. \textbf{Top Panel:} Normalized branching ratio $\overline{\mathcal{B}}$ as a function of cavity frequency $\omega_c$, classical (CM, blue circles) and quantum (QM, red squares). The on-resonance enhancement is observed near $\omega^{*}_{c_{on}}$ (purple dashed line), consistent with resonant coupling for $V_1^{\ddagger}$. Two off-resonant frequencies, $\omega^{l}_{c_{off}}$ (green) and $\omega^{r}_{c_{off}}$ (orange), are marked for comparison. \textbf{Bottom Panel:} Time-resolved product-domain probabilities $\mathcal{P}_D(t)$ for the marked frequencies in the top panel. The $\mathcal{P}_{P_1}(t)$ (solid) and $\mathcal{P}_{P_2}(t)$ (dotted), are obtained from CM (blue) and QM (red) dynamics at the off-resonant-left, on-resonant, and off-resonant-right cavity frequencies (left to right panels, color-coded borders matching the corresponding vertical line in the top panel). The enhanced branching ratio at $\omega^{*}_{c_{on}}$ correlates with a larger long-time population of $\mathcal{P}_{P_1}(t)$ relative to $\mathcal{P}_{P_2}(t)$, and CM and QM dynamics agree closely across all three frequencies. }
    \label{fig:domainprob_withtime}
    \alttext{A line graph of normalized branching ratio against cavity frequency above three time-resolved probability plots. The ratio peaks sharply near 3000 cm$^{-1}$, where the deeper-well population grows well above the other product, unlike at the two off-resonant frequencies.}
    \end{figure}

\section{Zero photon case}
\label{sec:zerophoton}

In the main manuscript, two different initial cavity states are considered. The cavity state used in Eq.~(5) in main,
\begin{equation*}
\phi(q_c) =
\left(\frac{\omega_c}{\pi\hbar}\right)^{1/4}
\exp\left(-\frac{\omega_c q_c^2}{2\hbar}\right),
\end{equation*}
corresponds to the ground state of the bare undisplaced cavity oscillator. For nonzero light-matter coupling, this state is not centered at the equilibrium position of the cavity coordinate in the coupled
Pauli-Fierz Hamiltonian. Consequently, its representation in the coupled light-matter system contains virtual photons.

A zero-photon cavity preparation can instead be constructed by displacing the cavity coordinate according to
\begin{equation*}
\phi(q_c;\lambda_c) =
\left(\frac{\omega_c}{\pi\hbar}\right)^{1/4} \exp\left[ -\frac{\omega_c(q_c-q_{c0})^2}{2\hbar} \right],
\end{equation*}
where \begin{equation*} q_{c0} = -\frac{\lambda_c}{\omega_c}\mu .
\end{equation*}
This state, given as Eq.~(6) in the main manuscript, corresponds to the displaced cavity ground state and removes the initial virtual-photon contribution. The subsequent dynamics in both cases are propagated using
the same fully coupled Pauli-Fierz Hamiltonian.
    \begin{figure}[H]
    \centering  
    \includegraphics[width=0.8\textwidth]{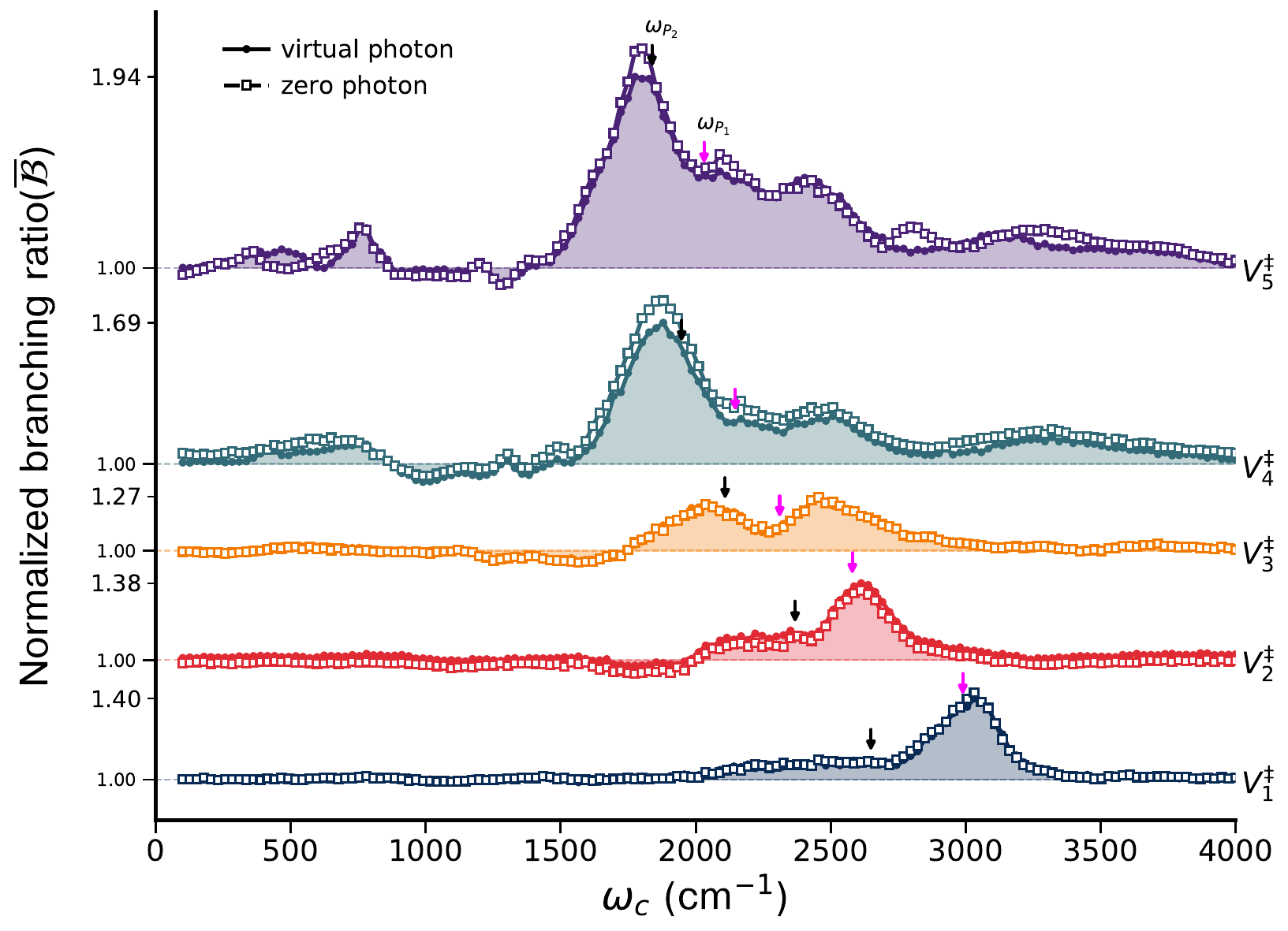}
    \caption{Branching ratios for the PTSB system under $x$-mode cavity coupling compared for the virtual and zero photon cases for five representative $V_n^{\ddagger}$.}
    \label{fig:ZPVP}
    \alttext{Stacked line graphs of normalized branching ratio against cavity frequency for virtual-photon and zero-photon initial cavity states, at five barrier parameter values. The two sets of curves overlap closely at every barrier value, differing only slightly in the peak height for the low barrier parameter values.}
    \end{figure}

To assess the sensitivity of the cavity-induced product selectivity to this choice of initial cavity preparation, we compare the normalized branching ratios obtained using the virtual-photon and zero-photon initial states. Fig. \ref{fig:ZPVP} shows this comparison for the five representative cases of $V_n^{\ddagger}$ shown in main. In both cases, the two initial-state preparations produce closely similar frequency-dependent branching-ratio profiles. In particular, associated frequency-dependent qualitative features in branching ratio are retained in the zero-photon calculations. Thus, frequency-dependent cavity-induced modification of the product branching ratio is not sensitive to whether the initial cavity state is taken to be virtual photon case of Eq.~(5) or the displaced zero-photon state of Eq.~(6).

    \begin{figure}[H]
    \centering
    \includegraphics[width=0.8\textwidth]{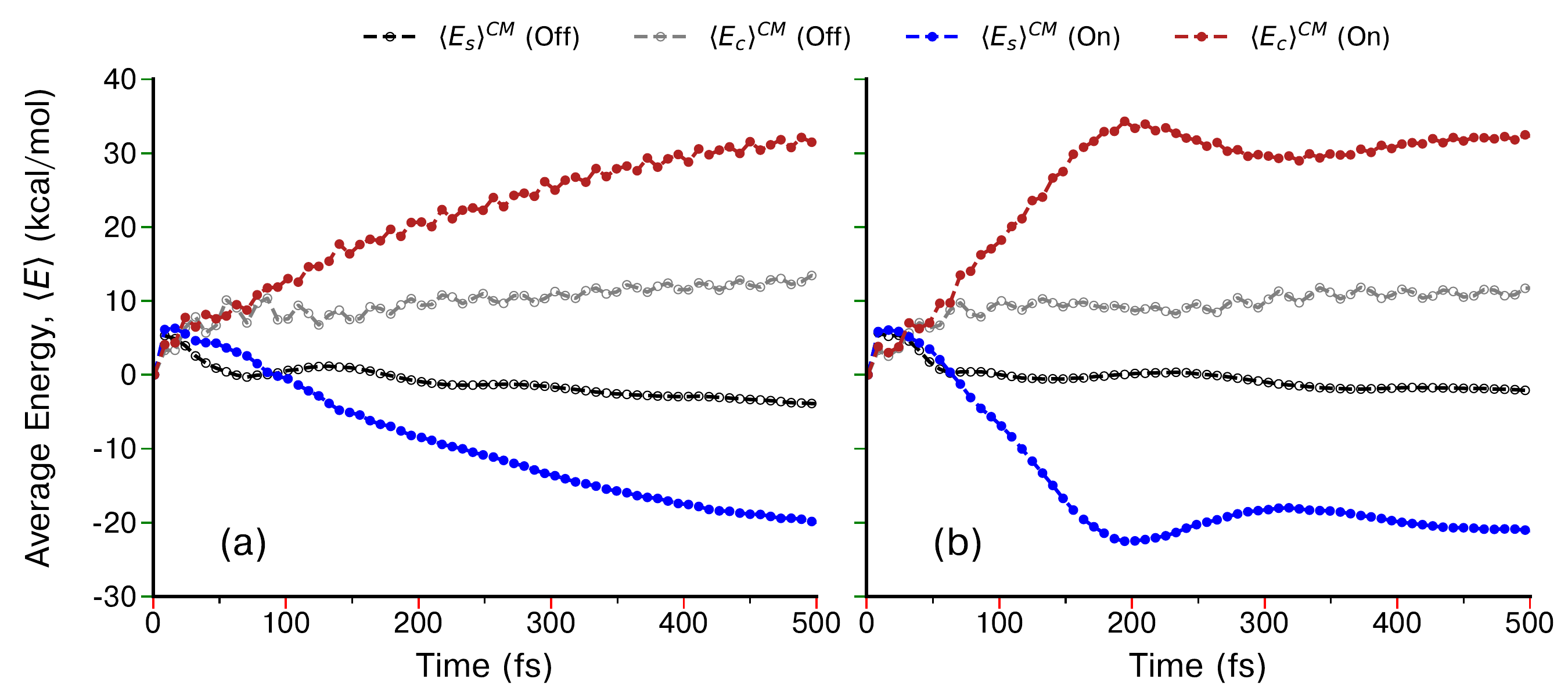}
    \caption{Average classical energies (indicated in the legend) for zero photon case are shown as a function of time for  (a) $V^{\ddagger}_{2}$ with $\omega_{c}=1200$ cm$^{-1}$ (off resonance), $\omega_{c}=\omega_{P_{1}}^{x} = 2586$ cm$^{-1}$ (on resonance) and (b) $V^{\ddagger}_{4}$ with $\omega_{c}=1071$ cm$^{-1}$ (off resonance),  $\omega_{c}=\omega_{P_{2}}^{x} = 1871$ cm$^{-1}$ (on resonance).}
    \label{fig:avgECM_V2V4ZP}
    \alttext{Two line graphs of average classical energy in kcal/mol against time up to 500 fs. Cavity energy rises and system energy falls in both panels; the on-resonance pair separates much further and sooner than the off-resonance pair.}
    \end{figure}

In addition to the branching-ratio comparison, we also examined the
time-dependent average energy exchange between the system and cavity  for the zero-photon initial cavity state Fig.\ref{fig:avgECM_V2V4ZP}. The classical ensemble-averaged energies defined in Sec.~\ref{sec:avgE} allows us to directly compare the cavity-system energy-exchange dynamics obtained from the zero-photon initial state with those discussed in the main manuscript for the virtual-photon initial state.

These results indicate that the observed selectivity modulation is not a consequence of initializing the cavity with virtual photons. We therefore use the virtual-photon initial state of Eq.~(5) for the main calculations and provide the zero-photon comparison here as a check of the robustness of the reported results.

\section{Computing average energy components from the Pauli-Fierz Hamiltonian}
\label{sec:avgE}

To analyze the energy exchange between different degrees of freedom, we
decompose the total light--matter Hamiltonian into system, cavity,
interaction, and dipole self-energy (DSE) contributions. Thus we can write 
\begin{equation*}
H = H_{\mathrm{S}} + H_{\mathrm{C}} +
        H_{\mathrm{int}}+ H_{\mathrm{DSE}} 
\end{equation*}
The matter (system) Hamiltonian describing the nuclear motion on the
PTSB potential energy surface is given by
\begin{equation*}
H_{\mathrm{S}} = \frac{p_x^{\,2}}{2m} +\frac{p_y^{\,2}}{2m} + U(x,y)
\end{equation*}
where $U(x,y)$ is the  two-dimensional PTSB potential given above. The Hamiltonian of the single mode cavity is given by
\begin{equation*}
H_{\mathrm{C}} =\frac{1}{2}\left(p_c^{\,2}+\omega_c^{2}q_c^{\,2}\right)
\end{equation*}
with $\omega_c$ being the cavity frequency.
The light-matter interaction term arises from the bilinear
coupling between the cavity $x$-coordinate and the molecular dipole
\begin{equation*}
H_{\mathrm{int}} = \lambda_c \, \omega_c \, q_c \, \mu(x),
\end{equation*}
where $\lambda_c$ is the light-matter coupling strength and $\mu(x)$ is the dipole function defined in the main text.
The energy associated with the dipole self-energy (DSE) term is given by
\begin{equation*}
H_{\mathrm{DSE}}=\frac{1}{2} \lambda_c^{2} \mu(x)^{2}
\end{equation*}
For the initial state of interest we time-evolve using the split-operator formalism to obtain the state $\Psi(t)$ and compute the time-dependent average energy components as
\begin{align*}
E^{QM}_{\mathrm{S}}(t)
&=
\langle \Psi(t) | H_{\mathrm{S}} | \Psi(t) \rangle, \\
 E^{QM}_{\mathrm{C}}(t) 
&=
\langle \Psi(t) | H_{\mathrm{C}} | \Psi(t) \rangle, \\
 E^{QM}_{\mathrm{int}}(t) 
&=
\langle \Psi(t) | H_{\mathrm{int}} | \Psi(t) \rangle, \\
 E^{QM}_{\mathrm{DSE}}(t) 
&=
\langle \Psi(t) | H_{\mathrm{DSE}} | \Psi(t) \rangle .
\end{align*}

The corresponding classical ensemble-averaged energies of each component are computed as 
\begin{align*}
 E^{CM}_{\mathrm{S}}(t) &= \frac{1}{N_{\mathrm{traj}}} \sum_{n=1}^{N_{\mathrm{traj}}} H^{(n)}_{\mathrm{S}}(x,y,p_{x},p_{y};t) \\
 E^{CM}_{\mathrm{C}}(t) & = \frac{1}{N_{\mathrm{traj}}} \sum_{n=1}^{N_{\mathrm{traj}}} H^{(n)}_{\mathrm{C}}(q_{c},p_{c};t) \\
 E^{CM}_{\mathrm{DSE}}(t) &= \frac{1}{N_{\mathrm{traj}}} \sum_{n=1}^{N_{\mathrm{traj}}} H^{(n)}_{\mathrm{DSE}}(x,q_{c};t) \\
 E^{CM}_{\mathrm{int}}(t) &= \frac{1}{N_{\mathrm{traj}}} \sum_{n=1}^{N_{\mathrm{traj}}} H^{(n)}_{\mathrm{int}}(x;t)
\end{align*}
where $N_{traj}$ is total ensemble size of trajectories used in the classical simulation.
The total average energy 
\begin{equation*}
 E_{\mathrm{tot}}  =
 E_{\mathrm{S}}  +
 E_{\mathrm{C}}  +
 E_{\mathrm{int}} +
 E_{\mathrm{DSE}}
\end{equation*}
is conserved in both quantum and classical cases.

\section{Methodology}
\label{sec:Method}
\subsection{Quantum dynamics}
\label{appenqm}
Starting with the initial polarition wavepacket, we solve the corresponding time-dependent Schr\"{o}dinger equation on a grid using the well-established split-operator method\cite{feit1983solution} involving the short time propagator
\begin{equation*}
    \hat{U}(\Delta t) = \exp\left(-i\frac{\Delta t}{2\hbar}\hat{V}\right) \exp \left(-i\frac{\Delta t}{\hbar}\hat{T} \right) \exp \left(-i\frac{\Delta t}{2\hbar}\hat{V}\right) \label{split_opr}
\end{equation*}
with $\hat{T}$ and $\hat{V}$ being the kinetic and potential energy operators respectively. The value of $\hbar$ is taken as unity in atomic units. Initial states are time-evolved over the timescale of interest with a step size of $\Delta t = 4$ a.u to ensure convergence. We chose a $256 \times 256$ spatial grid. The range of spatial coordinates for the molecular $x$, $y$ modes and the cavity mode are $[-1.5, 1.5]$, $[-1.5, 1.5]$, and $[-300, 300]$ au respectively.

\subsection{Classical dynamics}
\label{appencm}

The classical analog of the quantum wave packet is a Gaussian probability distribution $\rho(\mathbf{Q},\mathbf{P})$ with the same widths in position and momentum. Here, $\mathbf{Q}=(x,y,q_{c})$ and the corresponding conjugate momenta are denoted as $\mathbf{P}=(p_{x}, p_{y}, p_{c})$. The time evolution of this distribution is given by\cite{smdwksjcp2022}
\begin{equation*}
\begin{split}
    \rho_{cl}(\mathbf{Q},\mathbf{P},t) & = \int\int d\mathbf{Q}^{\prime}d\mathbf{P}^{\prime} \delta[\mathbf{Q}-\mathbf{Q}_{t}(\mathbf{P}^{\prime},\mathbf{Q}^{\prime})] \\ & \times \delta[\mathbf{P}-\mathbf{P}_{t}(\mathbf{P}^{\prime},\mathbf{Q}^{\prime})] \, \rho_{cl}(\mathbf{Q}^{\prime},\mathbf{P}^{\prime},0)  \label{rho_cl}
\end{split}
\end{equation*}
where $\mathbf{Q}_{t}(\mathbf{P}^{\prime},\mathbf{Q}^{\prime})$ and $\mathbf{P}_{t}(\mathbf{P}^{\prime},\mathbf{Q}^{\prime})$ represent the time-evolved phase-space coordinates as functions of the initial conditions. For our initial polaritonic wavepacket, the corresponding classical distribution $\rho_{cl}(\mathbf{Q},\mathbf{P},0) = \rho_{cl}(\mathbf{r},\mathbf{p},0) \rho_{cl}(q_{c},p_{c},0) \label{rho_cl_tot_0}$ is a Wigner distribution with $\mathbf{r}=(x,y)$ and $\mathbf{p}=(p_{x},p_{y})$, respectively centered at $(x_{0},p_{x_{0}}=0,y_{0},p_{y_{0}}=0,q_{c_{0}}=0,p_{c_{0}}=0)$ in the phase space.
The form of $\rho_{cl}(\mathbf{r},\mathbf{p},0)  \equiv \rho_{cl}(x,p_{x},0)\rho_{cl}(y,p_{y},0)$ is given by the expression
with
\begin{eqnarray}
        \rho_{cl}(x,p_{x},0) & = & \left(\frac{1}{\pi \sigma_{x} \sigma_{p_{x}}} \right)  \exp \left[-\frac{(x-x_{0})^{2}}{\sigma^{2}_{x}} - \frac{p_{x}^{2}}{\sigma^{2}_{p_{x}}}\right] \nonumber \\
        \rho_{cl}(y,p_{y},0)  &=& \left(\frac{1}{\pi \sigma_{y} \sigma_{p_{y}}} \right)  \exp \left[-\frac{(y-y_{0})^{2}}{\sigma^{2}_{y}} - \frac{p_{y}^{2}}{\sigma^{2}_{p_{y}}}\right] \nonumber
\end{eqnarray}
In the above, the position and momentum widths are denoted by $\sigma_{i} = (\hbar/m\omega^{i}_{\mathrm{R}})^{1/2}$ and $\sigma_{p_{i}} = (\hbar m\omega^{i}_{\mathrm{R}})^{1/2}$ respectively, with $(i=x,y)$. Similarly, for the cavity mode we take
\begin{equation}
    \rho_{cl}(q_{c},p_{c},0) = \left(\frac{1}{\pi \sigma_{q_{c}} \sigma_{p_{c}}} \right)  \exp \left[-\frac{q_{c}^{2}}{\sigma^{2}_{q_{c}}} - \frac{p_{c}^{2}}{\sigma^{2}_{p_{c}}}\right] \label{wignervirtualphoton}
\end{equation}
However, the above corresponds to the ``virtual photon"\cite{mondal2023phase} scenario in the optical cavity. To ensure the ``zero photon" condition, the corresponding Wigner distribution of the initial cavity state is taken to be
\begin{equation}
    \rho_{cl}(q_{c},p_{c},0) = \left(\frac{1}{\pi \sigma_{q_{c}} \sigma_{p_{c}}} \right)  \exp \left[-\frac{\left(q_{c}-q_{c_{0}}\right)^{2}}{\sigma^{2}_{q_{c}}} - \frac{p_{c}^{2}}{\sigma^{2}_{p_{c}}}\right] \label{wignerzerophoton}
\end{equation}
where, $q_{c_{0}}=-\frac{\lambda_{c}}{\omega_{c}}\mu$.  To compute classical domain probability, we initiate an ensemble of initial conditions sampled according to the initial density $\rho_{cl}(\mathbf{Q},\mathbf{P},0)$ and integrate their equation of motion forward in time using a fourth-order Runge–Kutta method.

\section{References}
\phantomsection
\label{sec:refs}

\vspace{-1.4cm}
\renewcommand{\refname}{}
\bibliographystyle{wiley-chemistry}
\bibliography{SI_PTSBVSC}